\documentclass[11pt,a4paper]{article}
\pdfoutput=1
\usepackage{jheppub}
\usepackage{extarrows}
\usepackage{caption}
\allowdisplaybreaks
\def\bea{\begin{eqnarray}}
\def\eea{\end{eqnarray}}
 \def\be{\begin{equation}}
\def\ee{\end{equation}}
\def\nn{\nonumber}

\usepackage[T1]{fontenc}

\title{Phenomenological study of neutrino mass, dark matter and baryogenesis within the framework of minimal extended seesaw}
  \author[a]{Pritam Das,}
  \author[b]{Mrinal Kumar Das}
  \author[c]{and Najimuddin Khan}
  \affiliation[a,b]{ Department of Physics, Tezpur University, Assam-784028, India}
\affiliation[c]{Centre for High Energy Physics, Indian Institute of Science,
	C. V. Raman Avenue, Bangalore 560012, India}
\affiliation[c]{School of Physical Sciences, Indian Association for the Cultivation of Science
	2A $\&$ 2B, Raja S.C. Mullick Road, Kolkata 700032, India}
  \emailAdd{pritam@tezu.ernet.in}
  \emailAdd{mkdas@tezu.ernet.in}
  \emailAdd{psnk2235@iacs.res.in}
  \abstract{We study a model of neutrino and dark matter within the framework of a minimal extended seesaw.
  	This framework is based on $A_4$ flavor symmetry along with the discrete $Z_4$ symmetry to stabilize the dark matter and construct desired mass matrices for neutrino mass.
  	We use a non-trivial Dirac mass matrix with broken $\mu - \tau$ symmetry to generate the leptonic mixing. A non-degenerate mass structure for right-handed neutrinos is considered to verify the observed baryon asymmetry of the Universe via the mechanism of thermal Leptogenesis. The scalar sector is also studied in great detail for a multi-Higgs doublet scenario, considering the lightest $Z_4$-odd as a viable dark matter candidate. A significant impact on the region of DM parameter space, as well as in the fermionic sector, are found in the presence of extra scalar particles. }

\arxivnumber{1911.07243}
\begin{document}
\maketitle
\flushbottom 
\section{Introduction}
The discovery of the Higgs boson ~\cite{Chatrchyan:2012xdj,Aad:2012tfa} at the Large Hadron Collider (LHC) confirms the mechanism of generation of the particle masses via the electroweak (EW) symmetry breaking. The Standard Model (SM) is a very successful theory to date, and it explains most of the experimental data. However, it is now known to be incomplete as it does not answer a few questions, like, non-zero neutrino masses,
baryon-antibaryon asymmetry in Nature, the mysterious nature of dark matter and dark energy, etc. Moreover, SM does not incorporate the theory of gravitation. Also, it is plagued with its theoretical problems such as the hierarchy problem related to the mass of the Higgs, mass hierarchy, and mixing patterns in leptonic and quark sectors, etc. The recent LHC Higgs signal strength data ~\cite{Sirunyan:2017khh, Sirunyan:2018koj} also suggests that one can have rooms for the new physics beyond SM.

The astrophysical observations like; gravitational lensing effects in Bullet cluster, anomalies in the galactic rotation
curves, etc., have been confirmed the existence of dark matter in the Universe. Dark matter is charge-less, so it cannot observe through its interactions with photons. The satellite-based experiments~\cite{Freese:2008cz, Persic:1995ru,Bennett:2012zja} such as Planck and Wilkinson Microwave Anisotropy Probe
(WMAP)  have measured the Cosmic Microwave Background
Radiation (CMBR) of the Universe with unprecedented accuracy. They have suggested that
the Universe consists of about 4$\%$ ordinary matter, 27$\%$ dark matter and the rest
69$\%$ is mysterious unknown energy called dark energy which is assumed to be the
cause of the accelerated expansion of the Universe. The SM of particle physics fails to provide a viable
dark matter candidate. Thus,
new physics beyond the SM is required to explain the observed presence of the dark matter.
The astrophysical and cosmological data so far can only tell us how much dark matter is
there in the Universe, i.e., the total mass density.

Although we are convinced
that dark matter exists, still, there is no consensus about its composition as it does not interact electromagnetically or strongly. The possibilities incorporate the dense-baryonic matter and non-baryonic matter. The MACHO and the EROS \cite{Tisserand:2006zx, Alcock:1998fx} collaborations conclude that
dense-baryonic matter, i.e, Massive Compact Halo Objects (MACHOs), black holes,  very faint stars, white dwarfs, non-luminous objects like planets
could add a few percent to the known mass discrepancy in the Galaxy halo observed in
galactic rotation curves. The non-baryonic dark matter components~\cite{Roszkowski:2017nbc} can be grouped into three categories based on their production mechanism and velocities, namely hot dark matter (HDM), warm dark matter
(WDM) and cold dark matter (CDM).
A simplified scheme in this regard is to introduce `weakly interacting massive particles' (WIMP) protected by a
discrete symmetry that ensures the stability of these particles. 
Various established options {\it viz.} extra $Z_n$-odd ($n\ge$2, is an integer) scalar, fermion, and combined of them with various multiplets, e.g., singlets, doublets, triplets, quadruplets, etc. have been studied to explain the dark matter phenomenology. One can see the recent review article on the dark matter~\cite{Hambye:2009pw, Bernal:2017kxu, Kahlhoefer:2017dnp, Tanabashi:2018oca, Magana:2012ph, Khan:2017xyh} and the references therein. 
The DM particles may have different masses: massive
gravitons $\mathcal{O}(10^{-19})$ GeV~\cite{Dubovsky:2004ud}, axions~\cite{Holman:1982tb} $\mathcal{O}(10^{-5})$ GeV, sterile neutrino~\cite{Dodelson:1993je} $\mathcal{O}(10^{-6})$ GeV and the point like WIMPs candidate having mass range $20$ GeV $-$ 340 TeV~\cite{Leane:2018kjk,Griest:1989wd, Hisano:2006nn}.
The upper bound on mass can be stretched up to 1 PeV for the composite dark matter candidate upper mass bound for the~\cite{Smirnov:2019ngs}. Keeping these bounds in mind, we study a point like WIMP dark matter candidate with a mass range in between 40$-$1000 GeV (can be reached upto $\mathcal{O}(100)$ TeV depending on the parameter space). It is also to be noted that one can ignore such upper bounds (coming from s-wave unitarity) may be varied here as the relic density mostly obtained by the $t,u$-annihilation and co-annihilation channels.

The other important experimental observation that necessitates the extension of the SM is the phenomenon of neutrino oscillation. The solar, atmospheric, reactor and accelerator neutrino oscillation experiments \cite{Abe:2016nxk, An:2012eh, Abe:2011fz} have shown that the three flavor neutrinos mix among themselves and they have a tiny mass but not zero, unlike as predicted by the SM. However, to date, the absolute mass of the individual neutrinos are not yet known. Even though, we get the sum of the all neutrino mass eigenvalues ($\sum_{i}<0.117$ eV \cite{Choudhury:2018byy}, with $i=1,2,3$) as the oscillation experiments are sensitive to mass square difference ($\Delta m_{ij}^2=m_i^2-m_j^2$). Apart from three-flavor oscillation, past results from Liquid Scintillation Neutrino Detector (LSND)~\cite{Athanassopoulos:1996ds} have found some anomalous results regarding neutrino mass. They have predicted a mass squared difference of $\sim \mathcal{O}(1)~\text{eV}^2$. This result contradicts the three neutrino theory, and we need to add another flavor of neutrino into the picture to explain the situation. Hence, the concept of a sterile neutrino is imposed in the neutrino picture. Later MiniBooNE~\cite{Aguilar-Arevalo:2018gpe} also confirmed the presence of this extra flavor of neutrino with $6.0\sigma$ confidence level whose mass roam around $\mathcal{O}(1)$ eV.

Along with these experimental signatures, various cosmological observations \cite{Hamann:2010bk, Izotov:2010ca} also support the presence of this extra flavor of massive neutrino. However, there is no concrete evidence on the number of generations of sterile neutrinos. Sterile neutrinos are added into the picture as a  right-handed (RH) particle, such that bare mass terms are allowed by all symmetries. Among the various beyond standard model scenarios that are proposed in the literature to explain the small neutrino masses, the most popular one is the seesaw mechanism. 

The seesaw mechanism is based on the assumption of the lepton number violated at a very high energy scale by some heavier particles. Tree level exchange of these heavy particles would give rise to the lepton number violating dimension-5 Weinberg operator $\Lambda^{-1} LLHH $\cite{weinberg}, which results in small neutrino masses once the EW symmetry is broken. 
It is clear that, to get a neutrino mass of the sub-eV scale, one has to take the new particles to be extremely heavy (right-handed neutrino) or else take the new couplings to be extremely small. To accommodate sterile neutrino along with the active neutrinos under the same roof, one can use the canonical type-I seesaw. The right-handed (RH) particles in type-I seesaw could have adjusted themselves as sterile neutrino simultaneously, giving rise to the active neutrino mass, if their masses lies in the eV-range. However, these kinds of possibilities are ruled out due to tiny active neutrino mass.
These difficulties are resolved by extending the type-I seesaw along with three additional RH heavy neutrinos, and a fermion singlet is popularly known as Minimal Extension Seesaw (MES)\cite{Zhang:2011vh, Barry:2011wb, Nath:2016mts} to study active neutrino masses along with the sterile mass in a single framework.

Considering various established suggestions regarding the evolution of our Universe, it is confirmed that, at the very beginning, there were equal numbers of matter and corresponding anti-matter. However, in the current scenario, there is an asymmetry in the baryon number is observed, and the scenario can be explained by the process, popularly known as baryogenesis. Numerical definition for baryon asymmetry at current date reads as \cite{Davidson:2008bu},
$
Y_{\Delta B}\big(\equiv \frac{n_B-n_{\overline{B}}}{s}\big)=(8.75\pm0.23)\times 10^{-11}$.
The SM does not have enough ingredient (Shakarov conditions) to explain this asymmetry. As seesaw demands lepton number violation, eventually new CP-violating phases in the neutrino Yukawa interactions are generated. It can be assumed that heavy singlet neutrinos decay out of equilibrium are able to produce this asymmetry. Thus, all three Sakharov conditions are satisfied naturally. Hence leptogenesis becomes an integral part of the seesaw framework. In this work, along with tiny neutrino masses, we will establish baryon asymmetry of the Universe produced via the mechanism of thermal leptogenesis under the MES framework.

We study this work in two sectors: the fermion and scalar sector. In the fermion sector, we have chosen the MES framework, where, along with the active neutrino mass generation, the validity of baryogenesis is checked in the presence of a heavy flavor of sterile neutrino. Sterile phenomenology with the active neutrinos has already been discussed in our previous work \cite{Das:2018qyt}. Within the scalar sector, we consider two additional Higgs doublets along with the SM Higgs doublet, where one additional Higgs doublet acquires VEV. At the same time, another one remains VEVless due to an additional $Z_4$ symmetry. The extension of scalar sector considering three Higgs doublets is quite popular in literature~\cite{Aranda:2012bv, Ivanov:2012ry, Chakrabarty:2015kmt,Pramanick:2015qga,Sokolowska:2017adz, Keus:2014jha}. The lightest neutral $Z_4$-odd Higgs doublet does not decay. Hence, it behaves as a potential candidate for dark matter. The other two scalar doublets are playing a crucial role in explaining the neutrino mass, mixing angles, and baryon asymmetry via leptogenesis. In the fermionic sector, they appear in the Dirac Lagrangian to give mass to the active neutrinos. In contrast, in the scalar sector, they influence the DM phenomenology and the allowed parameter spaces. All the theoretical and experimental constraints such as absolute stability, unitarity, EW precision, LHC Higgs signal strength, and dark matter density and direct detection are discussed in detail.

This paper is organized as follows. In section~\ref{sec2} brief review of the model is presented. We discuss the $A_4$ symmetry along with an additional $Z_4$ discrete symmetry to construct the model and generation of the mass matrices in the scalar and leptonic sector. We keep the section \ref{s3} and its subsections for presenting various bound on the scalar as well as fermion sector. Numerical
analysis of the dark matter and neutrino-sector along with the BAU results, are presented subsequently in section \ref{s4}. Finally, we conclude the summary of our work in section \ref{s5}.
\section{Sturcture of the model}\label{sec2}
 Discrete flavor symmetries like $A_4,S_4$\cite{Babu:2009fd, Ma:2009wi, Altarelli:2005yp, Mukherjee:2017pzq} along with $Z_n$ (n$\ge2$ is always an integer) are considered as intrinsic part of model building in particle physics. 
 $A_4$ being the discrete symmetry group of rotation with a tetrahedron invariant, it consists of 12 elements and 4 irreducible representation denoted by $\bf{1},\bf{1^{\prime}},\bf{1^{\prime\prime}} $ and $\bf{3}$. A brief discussion on $A_4$ and its product rules are carried out in appendix \ref{a4p}.
 We display the particle content and their charges within our model in table ~\ref{tab1}. The left-handed (LH) lepton doublet $l$ to transform as $A_4$ triplet, whereas right-handed (RH) charged leptons ($e^c,\mu^c,\tau^c$) transform as 1,$1^{\prime\prime}$ and $1^{\prime}$ respectively. The triplet flavons $\zeta, \varphi$ and two singlets $\xi$ and $\xi^{\prime}$  break the $A_4$ flavor symmetry by acquiring VEVs at large scale in the suitable directions\footnote{The chosen VEV alignments of the triplet flavons are obtained by minimizing the potential, can be found in the appendix of \cite{Das:2018qyt}. }. The Higgs doublets are assumed to be transformed as singlet under $A_4$. Additional $Z_4$ charges are assigned for the individual particles as per the interaction terms demands to restricts the non-desired terms.
 
 The $SU(2)$ doublet Higgs ($\phi_3$) along with $\xi^{\prime} $ and $\nu_{R3}$ are odd under $Z_4$. However, the mass scale of the flavon $\xi^{\prime}$ and the RH particle $\nu_{R3}$ are too heavy in comparison to the SU(2) doublet, $\phi_3$. As $\phi_3$ is a $Z_4$ odd field,  it doesn't couple with any SM fields, and so the lightest neutral doublet does not decay to any SM particle directly. Indirect detection experiments constrain DM decay lifetime to be larger than $10^{27} -10^{28} ~s$ \cite{Cohen:2016uyg}, in the meanwhile, the dark matter can decay to the SM particles through $\xi^{\prime}$, $\nu_{R1}$ and $\nu_{R2}$ via 7,8-body decay processes, which is heavily suppressed by the propagator masses and decay width is almost zero. The lifetime, in this case, is much greater than $10^{60} ~s$. Hence, the lightest neutral particle of $\phi_3$ is stable, and it can serve as a viable cold-WIMP dark matter candidate. 
  	 \begin{table}[h!]
	\begin{tabular}{c|ccccc|cc|cccc|ccc|cc}
		\hline
		Particles & $l$ &$e_{R}$&$\mu_{R}$&$\tau_{R}$&$\phi_1$&$\phi_2$&$\phi_3$&$\zeta$&$\varphi$&$\xi$&$\xi^{\prime}$&$\nu_{R1}$&$\nu_{R2}$&$\nu_{R3}$&$S$&$\chi$\\
		\hline
		
		SU(2)&2&1&1&1&2&2&2&1&1&1&1&1&1&1&1&1\\
		\hline
		$A_4$&3&1&$1^{\prime\prime}$&$1^{\prime}$&1&1&1&3&3&1&$1^{\prime}$&1&$1^{\prime}$&1&$1^{\prime\prime}$&$1^{\prime}$\\
		\hline
		$Z_4$&1&1&1&1&1&i&-1&1&i&1&-1&1&-i&-1&i&-i\\
		\hline
		\hline		
	\end{tabular}
	\caption{Particle content and their charge assignments under SU(2), $A_4$ and $Z_4$ groups. The second block of the particle content ($l,e_R,\mu_R,\tau_R,\phi_1$) represents the left-handed lepton doublet, RH charged fermions and SM Higgs doublets respectively. $\phi_2$ and $\phi_3$ (inert) are the additional Higgs doublet. $\nu_{Ri}(i=1,2,3)$ and $S$ are RH neutrinos and chiral singlet. Rest of the particles ($\zeta,\varphi,\xi,\xi^{\prime},\chi$) are the additional flavons.}\label{tab1} 
\end{table}
\subsection{Scalars}
The doublet Higgs scalars in this model are conventionally expressed as\cite{Keus:2013hya},
\begin{equation}
\phi_1=\begin{pmatrix}
H_1^+\\ \frac{(H_1+iA_1)}{\sqrt{2}}\\
\end{pmatrix}\quad;\quad   \phi_2=\begin{pmatrix}
H_2^+\\ \frac{(H_2+iA_2)}{\sqrt{2}}\\
\end{pmatrix}\quad ; \quad   \phi_3=\begin{pmatrix}
H_{3}^+\\ \frac{(H_{3}+iA_{3})}{\sqrt{2}}\\
\end{pmatrix}.
\end{equation}
The kinetic part of the scalar is defined within SM paradigm as,
\begin{equation}
\mathcal{L}_{KE}=\sum_{i=1}^3(D_{\mu}\phi_{i})^{\dagger}(D_{\mu}\phi_{i}),
\end{equation}
where, $D_{\mu}$ stands for the covariant derivative. The scalar potential of the Lagrangian is written in two separate parts. Among the three Higgs doublets, one of them does not acquire any VEV, so it behaves as inert while the other two are SM type Higgs doublet and acquire VEV by EWSB. The scalar potential of the Lagrangian is defined as\cite{Deshpande:1977rw},
\begin{equation}
	\mathcal{V}_{\phi_1,\phi_2,\phi_3}=\big(V_{\phi_1+\phi_2}+V_{\phi_3}\big),
\end{equation}
where,
\allowdisplaybreaks
\begin{equation}
\begin{split}\label{12pot}
V_{\phi_1+\phi_2}&=\mu^2_{11}\phi_1^{\dagger}\phi_1+\mu^2_{22}\phi_2^{\dagger}\phi_2-\frac{\mu^2_{12}}{2}(\phi_1^{\dagger}\phi_2+\phi_2^{\dagger}\phi_1)\\
&+\kappa_1(\phi_1^{\dagger}\phi_1)^2+\kappa_2(\phi_2^{\dagger}\phi_2)^2+\kappa_3(\phi_1^{\dagger}\phi_1)(\phi_2^{\dagger}\phi_2)\\
&+\kappa_4(\phi_2^{\dagger}\phi_1)(\phi_1^{\dagger}\phi_2)+\frac{\kappa_5}{2}((\phi_1^{\dagger}\phi_2)^2+(\phi_2^{\dagger}\phi_1)^2),
\end{split}
\end{equation}
while potential for the inert Higgs is given as,
\allowdisplaybreaks
\begin{equation}
\begin{split}
V_{\phi_3}&=\mu^2_{33}\phi_3^{\dagger}\phi_3+\kappa_2^{DM}(\phi_3^{\dagger}\phi_3)^2+\kappa_3^{DM}((\phi_1^{\dagger}\phi_1)(\phi_3^{\dagger}\phi_3)+(\phi_2^{\dagger}\phi_2)(\phi_3^{\dagger}\phi_3))\nn\\
&+\kappa_4^{DM}((\phi_1^{\dagger}\phi_3)^{\dagger}(\phi_1^{\dagger}\phi_3)+(\phi_2^{\dagger}\phi_3)^{\dagger}(\phi_2^{\dagger}\phi_3))+\kappa_5^{DM}((\phi_1^{\dagger}\phi_3)^2+(\phi_2^{\dagger}\phi_3)^2+h.c.).
\end{split}
\end{equation}
In both the potentials, $\mu_{ij} (i=1,2)$, $\mu_{33}$ are the mass terms and $\kappa's$ are the scalar quartic couplings, responsible for mixing and masses of the physical scalar fields. The neutral CP-even fields of $\phi_1$ and $\phi_2$ get VEVs after electroweak symmetry breaking (EWSB), i.e., $H_1=h_1+v_1$ and $H_2=h_2+v_2$.
The minimization conditions for the potential are,
\begin{eqnarray}
\mu^2_{11}&=\mu_{12}^2\tan\beta-\frac{1}{2}v^2\big(2\kappa_1\cos^2\beta+\kappa_L\sin^2\beta\big),\\
\mu^2_{22}&=\mu_{12}^2\cot\beta-\frac{1}{2}v^2\big(2\kappa_2\sin^2\beta+\kappa_L\cos^2\beta\big).\nn
\end{eqnarray}
Here, $\kappa_L=(\kappa_1+\kappa_2+\kappa_3)$, $v=\sqrt{v_1^2+v_2^2}$ and $\beta=\tan^{-1}\big(\frac{v_2}{v_1}\big)$. It is to be noted that there is no minimum along with the directions of the scalar fields in $\phi_3$ doublet due to an additional $Z_4$ symmetry.

A $12\times12$ mass matrix is obtained after EWSB, which is composed of four $3\times3$ sub-matrices with bases $(H_1^+,H_2^+,H_3^+)$ , $(H_1^-,H_2^-,H_3^-)$ , $(h_1,h_2,H_3)$ and $(A_1,A_2,A_3)$. The inert fields in these mass matrices remain decoupled as they do not get any VEV. The other fields give rise to five physical mass eigenstates $(H^{\pm},h,H,A)$ after rotation with the mass basis. Three other mass-less Goldstone bosons $(G^{\pm},G^0)$ are also generated, which are eaten up by the $W^{\pm}$ and $Z$ bosons to give mass them mass. The mass eigenstates for the physical scalars within 2HD (first two Higgs doublets) scalar sector are given by \cite{Branco:2011iw},
\allowdisplaybreaks
\begin{eqnarray}
 M_{A}^2&=& \frac{\mu_{12}^2}{2\cos\beta \sin\beta}-\kappa_5v^2,\\
 M_{H^{\pm}}^2&=&  M_A^2+\frac{1}{2}v^2(\kappa_5-\kappa_4),\\
 M_h^2&=&  \frac{1}{4}v^2\sec(\alpha+\beta)[(6\kappa_1+\kappa_L)\cos\alpha\cos\beta\nn\\
&& +2(\kappa_1-\kappa_L)\cos\alpha\cos3\beta-\sin\alpha\sin\beta\, \{(6\kappa_2+\kappa_L)-(2\kappa_2+\kappa_L)\}],\\
 M_H^2&=&  \frac{1}{4}v^2\mathrm{cosec}(\alpha+\beta) \{2\cos\beta(2\kappa_1+\kappa_L+(2\kappa_1-\kappa_L)\cos2\beta\}\sin\beta\nn\\
&&+\cos\alpha \, \{(6\kappa_2+\kappa_L)\sin\beta+(-2\kappa_2+\kappa_L)\sin3\beta\},\\
\text{where,}\nn\\
\mu_{12}^2&=& \frac{1}{2}v^2[\kappa_1+\kappa_2+\kappa_L+\mathrm{cosec}(2\alpha+2\beta)\{2(\kappa_1-\kappa_2)\sin2\alpha+(\kappa_1+\kappa_2-\kappa_L)\sin(2\alpha-2\beta)\}]\sin2\beta.\nn
\end{eqnarray}
Inert scalar sector remain decoupled from the other two scalar sector. After EWSB four physical mass eigenstates ($H_3,A_3,H_3^{\pm}$) can be written as,
\begin{equation}
 \begin{split}
M_{H_3}^2=&\mu_{33}^2+\frac{1}{2}v^2\kappa_L^{DM},\\
M_{A_3}^2=&\mu_{33}^2+\frac{1}{2}v^2\kappa_S^{DM},\\
M_{H_3^{\pm}}^2=&\mu_{33}^2+\frac{1}{2}v^2\kappa_3^{DM},\\
 \end{split}
\end{equation}
where, $\kappa_{L,S}^{DM}=\kappa_3^{DM}+\kappa_4^{DM}\pm \kappa_5^{DM}$.
It is to be noted that the detailed study of the scalar potential and the interaction among the heavy scalar fields ($\varphi,\zeta,\xi,etc.$) are worked out in \cite{Das:2018qyt}, and the light scalar fields remains decoupled from these heavy scalar fields. 
These heavy scalar fields need to explain neutrino mass and mixing which we are going to discuss in the next subsection.
\subsection{Fermions}
 Minimal extended seesaw(MES) is realized in this work to construct active and sterile masses. There is a similar kind of framework named $\nu MSM$ \cite{Shaposhnikov:2006nn}, exists in literature, which is an extension of SM, where keV sterile neutrino mass is studied. However, MES is an extension of the canonical type-I seesaw, where sterile neutrino mass is realized with a broader mass range (eV to keV) than $\nu MSM$. We have focused our study with eV scaled sterile neutrino and three flavors of active neutrinos. In the MES scenario, along with the SM particle, three extra right-handed neutrinos, and one additional gauge singlet chiral field, $S$ is introduced. The Lagrangian of the neutrino mass terms for MES is given by\cite{Barry:2011wb},
\begin{equation}\label{mes1}
-\mathcal{L}_{\mathcal{M}}= \overline{\nu_{L}}M_{D}\nu_{R}+\frac{1}{2}\overline{\nu^{c}_{R}}M_{R}\nu_{R}+\overline{S^c}M_{S}\nu_{R}+h.c. .
 \end{equation}  
Here, $M_D$ and $M_R$ are $3\times3$ Dirac and Majorana mass matrices respectively whereas $M_S$ is a $1\times3$ matrix. Under MES framework, the active and sterile masses are realized as follows,
\begin{equation}\label{amass}
 m_{\nu}\simeq M_{D}M_{R}^{-1}M_{S}^T(M_{S}M_{R}^{-1}M_{S}^{T})^{-1}M_{S}(M_{R}^{-1})^{T}M_{D}^{T}-M_{D}M_{R}^{-1}M_{D}^{T}, 
\end{equation}
\begin{equation}\label{smass}
 m_{s}\simeq -M_{S}M_{R}^{-1}M_{S}^{T}.
\end{equation}
One can see that the second part of eq.~\eqref{amass} is the trivial type-I active neutrino mass formula. The modified active mass matrix is due to the presence of the non-squared sterile mass matrix $M_S$. The mass scale of $M_S$ being slightly higher than the $M_D$ scale, which is near to the EW scale, while $M_R$ is around $10^{13}$ GeV. A new physics scale ($\Lambda$) is imposed to achieve this MES structure. $A_4$ flavon symmetry has been broken in the model by the heavy scale, in order to generate the required neutrino mass, thus making the scale very heavy ($\sim10^{14}$ GeV).  

The leading order invariant Yukawa Lagrangian for the lepton sector is given by,
 \begin{equation}\label{lag}
\begin{split}
\mathcal{L} =& \frac{y_{e}}{\Lambda}(\overline{l} \phi_1 \zeta)_{1}e_{R}+\frac{y_{\mu}}{\Lambda}(\overline{l} \phi_1 \zeta)_{1^{\prime}}\mu_{R}+\frac{y_{\tau}}{\Lambda}(\overline{l} \phi_1 \zeta)_{1^{\prime\prime}}\tau_{R} \\
&+ \frac{y_{2}}{\Lambda}(\overline{l}\tilde{\phi_1}\zeta)_{1}\nu_{R1}+\frac{y_{2}}{\Lambda}(\overline{l}\tilde{\phi_1}\varphi)_{1^{\prime\prime}}\nu_{R2}+\frac{y_{3}}{\Lambda}(\overline{l}\tilde{\phi_2}\varphi)_{1}\nu_{R3}\\
& +\frac{1}{2}\lambda_{1}\xi\overline{\nu^{c}_{R1}}\nu_{R1}+\frac{1}{2}\lambda_{2}\xi^{\prime}\overline{\nu^{c}_{R2}}\nu_{R2}+\frac{1}{2}\lambda_{3}\xi\overline{\nu^{c}_{R3}}\nu_{R3}\\
&+ \frac{1}{2}\rho\chi\overline{S^{c}}\nu_{R1} .\\
\end{split}
\end{equation}
$\Lambda$ in the Lagrangian, represents the cut-off scale of the theory, $y_{\alpha,i}$, $\lambda_{i}$ (for $\alpha=e,\mu,\tau$ and $i=1,2,3$) and $\rho$ representing the Yukawa couplings for respective interactions and all Higgs doublets are transformed as $\tilde{\phi_i} = i\tau_{2}\phi_i^*$ (with $\tau_{2}$ being the second Pauli's spin matrix)  to keep the Lagrangian gauge invariant.
The scalar flavons involved in the Lagrangian acquire VEV along  $ \langle \zeta \rangle=(v_m,0,0), \langle\varphi\rangle=(v_m,v_m,v_m), 
\langle\xi\rangle=\langle\xi^{\prime}\rangle=v_m$ and $ \langle\chi\rangle=v_{\chi}$ by breaking the flavor symmetry, while $\langle \phi_i\rangle(i=1,2)$ get VEV ($v_i$) by breaking EWSB at electro-weak scale ($v_3=0$ due to additional $Z_4$ symmetry). The matrix structures obtained from equation \ref{lag} using the VEV of these flavons are discussed in appendix \ref{matrix}. We achieve the light neutrino mass matrix as well as the sterile mass using eq. \eqref{amass} and \eqref{smass} respectively. Complete matrix structures are shown in table \ref{tab:nh}. In the following section, we have presented detailed theoretical as well as experimental bounds on this model.
\section{Bounds on this models}\label{s3}
\subsection{Stability of the scalar potential}
The scalar potential should be bounded from the below in such a way that, even for large field values there is no other negative infinity that arises along any field space direction. The absolute stability condition for the potential [\ref{12pot}] are evaluated in terms of the quadratic coupling are as follows~\cite{Deshpande:1977rw},
\bea
\kappa_1>0, ~ \kappa_2>0,~ \kappa_2^{DM}>0, \kappa_{3,L,S}+2\sqrt{2\kappa_1\kappa_2}>0,&\nn\\
\kappa_{3,L,S}^{DM}+2\sqrt{\kappa_1\kappa_2^{DM}}>0,  \kappa_{3,L,S}^{DM}+2\sqrt{\kappa_2\kappa_1^{DM}}>0.\nn
\eea
\subsection{Unitarity bounds}
Unitarity bounds on the couplings are evaluated considering scalar-scalar, gauge boson-gauge boson, and
scalar-gauge boson scatterings \cite{Lee:1977eg}. In general, unitarity bounds are the couplings of the physical bases of the scalar potential. However, the couplings for the scalars are quite complicated, so we consider the couplings of the non-physical bases before EWSB. Then the S-matrix, which is expressed in terms of the non-physical fields, is transformed into an S-matrix for the physical
fields by making a unitary transformation \cite{Das:2014fea, Arhrib:2012ia, Kanemura:1993hm}. The unitarity of the S-matrix demands the absolute eigenvalues of the scattering matrix should be less than $8\pi$ up to a particular scale. In our potential, bounds come from the eigenvalues of the corresponding S-matrix are as follows,

\begin{minipage}[c]{0.50\textwidth}
\begin{equation}
\resizebox{0.98\hsize}{!}{ $
\begin{split}
& |\kappa_3\pm \kappa_4 |\leq8\pi,\\
& |\kappa_3\pm \kappa_5 |\leq 8\pi,\\
&|\kappa_3+2\kappa_4\pm 3\kappa_5|\leq 8\pi,\\
&\Big|\kappa_1+\kappa_2\pm \sqrt{(\kappa_1-\kappa_2)^2+\kappa_4} \Big|\leq 8\pi,\\
&\Big|3\kappa_1+3\kappa_2\pm \sqrt{9(\kappa_1-\kappa_2)^2+(2\kappa_3+\kappa_4)^2}\Big|\leq 8\pi,\\
&\Big| \kappa_1+\kappa_2\pm \sqrt{(\kappa_1-\kappa_2)^2+\kappa_5} \Big|\leq 8\pi,\\
&\Big| 3\kappa_2+3\kappa_2^{DM} \pm \sqrt{9(\kappa_2-\kappa_2^{DM})^2+(2\kappa_3^{DM}+\kappa_4^{DM})^2} \Big|\leq 8\pi,\\
&\Big| \kappa_2+\kappa_2^{DM}\pm \sqrt{(\kappa_2-\kappa_2^{DM})^2+2\kappa_5^{DM}} \Big|\leq 8\pi,\nn
\end{split}$}
\end{equation}
\end{minipage}
\hspace{-0.4cm}
\begin{minipage}[c]{0.50\textwidth}
\begin{equation*}
\resizebox{0.98\hsize}{!}{ $
\begin{split}
&|\kappa_3^{DM}\pm \kappa_4^{DM}|\leq 8\pi,\\
&|\kappa_3^{DM}\pm 2\kappa_5^{DM}|\leq 8\pi,\\
&|\kappa_3^{DM}+2\kappa_4^{DM}\pm 6\kappa_5^{DM}|\leq 8\pi,\\
&\Big|\kappa_1+\kappa_2^{DM}\pm \sqrt{(\kappa_1-\kappa_2^{DM})^2+\kappa_4^{DM}}\Big|\leq 8\pi,\\
&\Big|3\kappa_1+3\kappa_2^{DM}\pm \sqrt{9(\kappa_1-\kappa_2^{DM})^2+(2\kappa_3^{DM}+\kappa_4^{DM})^2} \Big| \leq 8\pi,\\
&\Big|\kappa_1+\kappa_2^{DM}\pm \sqrt{(\kappa_1-\kappa_2^{DM})^2+2\kappa_5^{DM}}\Big|\leq 8\pi,\\
&\Big|\kappa_2+\kappa_2^{DM}\pm \sqrt{(\kappa_2-\kappa_2^{DM})^2+\kappa_4^{DM}}\Big|\leq 8\pi.\nn
\end{split} $}
\end{equation*}
\end{minipage}
\subsection{Bounds from electroweak precision experiments}
When we consider a new physics contribution above the EW scale, the effect of the virtual particles in loops does contribute to the electroweak precision bounds through vacuum polarization correction. Bounds from electroweak precision experiments are added in new physics contributions via self-energy parameters $S,T,U$ \cite{Baak:2014ora}. The $S$ and $T$ parameters provide the new physics contributions to the neutral and the difference between neutral and charged weak currents, respectively. In contrast, the $U$ parameter is only sensitive to the mass and width of the W-boson; thus, in some cases, this parameter is neglected. In this model, inert scalars decouple from the other scalar fields. Contributions from first two doublet fields are \cite{Chakrabarty:2015kmt},
\bea
\Delta S_{2HD}&=&\frac{1}{\pi M_Z^2}\Big[\sin^2(\beta-\alpha)\mathcal{B}_{22}(M_Z^2,M_H^2,M_A^2)-\mathcal{B}_{22}(M_Z^2,M_{H^{\pm}}^2,M_{H^{\pm}}^2)\nn\\
&+&\cos^2(\beta-\alpha)\Big\{\mathcal{B}_{22}(M_Z^2,M_h^2,M_A^2)+\mathcal{B}_{22}(M_Z^2,M_Z^2M_H^2)-\mathcal{B}_{22}(M_Z^2,M_Z^2,M_h^2)\nn\nn\\
&-&M_Z^2\mathcal{B}_{0}(M_Z^2,M_Z^2,M_H^2)+M_Z^2\mathcal{B}_{0}(M_Z^2,M_Z^2,M_h^2)\Big\}\Big],\\
\Delta T_{2HD}&=&\frac{1}{16\pi M_W^2\sin^2_{\theta_W}}\Big[F(M^2_{H^{\pm}},M^2_A)+\sin^2(\beta-\alpha)\Big\{F(M^2_{H^{\pm}},M^2_H)-F(M^2_A,M^2_H) \}\nn\\
& +&\cos^2(\beta-\alpha)\Big\{F(M^2_{H^{\pm}},M^2_h)-F(M^2_A,M^2_h)+F(M^2_W,M^2_H)-F(M^2_W,M^2_h)\nn\\
&-&F(M^2_Z,M^2_H)+F(M^2_Z,M^2_h)+4M_Z^2\mathcal{\overline{B}}_{0}(M_Z^2,M_H^2,M_h^2)\nn\\
&-&M_Z^2\mathcal{\overline{B}}_{0}(M_W^2,M_H^2,M_h^2)\}\Big],\\
\Delta U_{2HD}&=&-S+\frac{1}{\pi M_Z^2}\Big[\mathcal{B}_{22}(M_W^2,M_A^2,M_{H^{\pm}}^2)-2\mathcal{B}_{22}(M_W^2,M^2_{H^{\pm}},M_{H^{\pm}}^2)\nn\\
&+&\sin^2(\beta-\alpha)\mathcal{B}_{22}(M_W^2,M_H^2,M_{H^{\pm}}^2)\nn\\
& +&\cos^2(\beta-\alpha)\Big\{\mathcal{B}_{22}(M_W^2,M_h^2,M_{H^{\pm}}^2)+\mathcal{B}_{22}(M_W^2,M_W^2,M_{H}^2)-\mathcal{B}_{22}(M_W^2,M_W^2,M_h^2)\nn\\
&-&M_W^2\mathcal{B}_{0}(M_W^2,M_W^2,M_{H}^2)+M_W^2\mathcal{B}_{0}(M_W^2,M_W^2,M_{H}^2)\Big\}\Big],
\eea
while the contributions from inert fields can be written as \cite{Arhrib:2012ia, Baak:2014ora},
\bea
\Delta S_{ID}&=&\frac{1}{2\pi}\Big[\frac{1}{6}\ln\frac{M_H^2}{M_{H^{\pm}}^2}-\frac{5}{36}+\frac{M_H^2M_A^2}{3(M_A^2-M_H^2)^2}+\frac{M_A^4(M_A^2-3M_H^2)}{6(M_A^2-M_H^2)^3}\ln \frac{M_A^2}{M_H^2} \Big],\\
\Delta T_{ID}&=&\frac{1}{32\pi^2\alpha v^2}\Big[F(M^2_{H^{\pm}},M^2_H)+F(M^2_{H^{\pm}},M_A^2)-F(M_A^2,M^2_H)\Big].
\eea
$\Delta U_{ID}$ is neglected in this case due to small mass differences $\Delta M_{H}=M_{A_3}-M_{H_3}$ and $\Delta M_{H^\pm}=M_{H_3^\pm}-M_{H_3}$ of the inert fields.
$F$ and $\mathcal{B}$'s are defined as,
\begin{eqnarray*}
F(x,y)&=&\frac{x+y}{2}-\frac{xy}{x-y}\ln (\frac{x}{y}), \quad \text{for}\quad x\ne y \quad\text{otherwise}\quad 0.\\
\mathcal{B}_{22}(q^2,m_1^2,m_2^2)&=&\frac{q^2}{24}[2\ln q^2+\ln(x_1x_2)+\{(x_1-x_2)^3-3(x_1^2-x_2^2)+3(x_1-x_2)\}\ln (x_1/x_2)\nn\\
&&-\{2(x_1-x_2)^2-8(x_1+x_2)+\frac{10}{3}\}-\{(x_1-x_2)^2-x(x_1+x_2)+1\}f(x_1,x_2)\nn\\
&&-6F(x_1,x_2)]  \xLongrightarrow{m_1=m_2}\frac{q^2}{24}[2\ln q^2+\ln x_1+(16x_1-\frac{10}{3})+(4x_1-1)G(x_1)],\\
\mathcal{B}_0(q^2,m_1^2,m_2^2)&=& 1+\frac{1}{2}[\frac{x_1+x_2}{x_1-x_2}-(x_1-x_2)]\ln(x_1/x_2)+\frac{1}{2}f(x_1,x_2)\xLongrightarrow{m_1=m_2}2-2y\arctan\frac{1}{y},\\
\mathcal{\overline{B}}_0(m_1^2,m_2^2,m_3^2)&=&\frac{m_1^2\ln m_1^2-m^2_3\ln m_3^2}{m_1^2-m_3^2}-\frac{m_1^2\ln m_1^2-m_2^2\ln m_2^2}{m_1^2-m_2^2},\\
\text{with},\quad &&x_i=m_i^2/q^2 \quad , \quad y=\sqrt{4x_1-1}\quad, \quad G(x_1)= -4y\arctan\frac{1}{y},\\ 
&&f(x_1,x_2)=-2\sqrt{\Upsilon}[\arctan(\frac{x_1-x_2+1}{\sqrt{\Upsilon}})-\arctan(\frac{x_1-x_2-1}{\sqrt{\Upsilon}})]\quad\text{for}\quad \Upsilon>0,\\
&&f(x_1,x_2)=\sqrt{-\Upsilon}\Big[\ln(\frac{x_1+x_2-1+\sqrt{-\Upsilon}}{x_1+x_2-1-\sqrt{-\Upsilon}})\Big]\quad\text{for}\quad \Upsilon<0,\\\text{and}\quad && f(x_1,x_2)=0 \quad \text{for}\quad  \Upsilon=0;\quad \text{where,} \quad \Upsilon=2(x_1+x_2)-x_1-x_2
^2-1.
\end{eqnarray*}
One can add these contributions to the SM as,
\begin{equation}
\begin{split}
S=S_{SM}+\Delta S_{IDM}+\Delta S_{2HDM},\\
T=T_{SM}+\Delta T_{IDM}+\Delta T_{2HDM},\\
U=U_{SM}+\Delta U_{IDM}+\Delta U_{2HDM}.\\
\end{split}
\end{equation}
We use the NNLO global electroweak fit results obtained by the Gfitter
group~\cite{Baak:2014ora}, $\Delta S_{IDM}+\Delta S_{2HDM}<0.05\pm0.11$, $T_{IDM}+\Delta T_{2HDM}<0.09\pm0.13$ and $\Delta U_{IDM}+\Delta U_{2HDM}<0.011\pm0.11$.
\subsection{LHC diphoton signal strength bounds}
At tree-level, the couplings of Higgs-like scalar $h$ to the fermions and gauge bosons in the presence of extra Higgs doublet ($\phi_2$) are modified due to the mixing. Loop induced decays will also have slight modification for the same reason. Hence, new contributions will be added to the signal strength. Using narrow width approximation, $\Gamma_h/M_h\rightarrow0$, the Higgs to diphoton strength is,
\begin{equation}
\mu_{\gamma\gamma}=\frac{\sigma(gg\rightarrow h\rightarrow \gamma\gamma)_{BSM}}{\sigma(gg\rightarrow h\rightarrow \gamma\gamma)_{SM}}\approx\frac{\sigma(gg\rightarrow h)_{BSM}}{\sigma(gg\rightarrow h)_{SM}}\frac{Br(h\rightarrow\gamma\gamma)_{BSM}}{Br(h\rightarrow\gamma\gamma)_{SM}}.
\end{equation}
In presence of an extra inert Higgs doublet $\phi_3$, the signal strength does not change, however, due to mixing of $\phi_1$ and $\phi_2$, $h$ to flavon-flavon ( or boson-boson) coupling become proportional to $\frac{\sin\alpha}{\cos\beta}(\text{or}\cos(\beta-\alpha))$. So, we may rewrite $\mu_{\gamma\gamma}$ as,
\begin{equation}
\mu_{\gamma\gamma}=\frac{\sin^2\alpha}{\cos^2\beta}\frac{\Gamma(h\rightarrow\gamma\gamma)_{BSM}}{\Gamma(h\rightarrow\gamma\gamma)_{SM}}\frac{\Gamma^{total}_{h,SM}}{\Gamma^{total}_{h,BSM}}.
\end{equation}
Apart from the SM Higgs $h$, if the masses for the extra physical Higgses are greater than $M_h/2$, then $\frac{\Gamma^{total}_{h,SM}}{\Gamma^{total}_{h,BSM}}\approx\big(\frac{\sin^2\alpha}{\cos^2\beta}\big)^{-1}$. Hence, the modified signal strength will be written as,
\begin{equation}
\mu_{\gamma\gamma}=\frac{\Gamma(h\rightarrow\gamma\gamma)_{BSM}}{\Gamma(h\rightarrow\gamma\gamma)_{SM}}.
\end{equation}
At one-loop level, the physical Higgs $H^{\pm}$ and $H_3^{\pm}$ add extra contribution to the decay width as,
\begin{equation}
\Gamma(h\rightarrow \gamma\gamma)_{BSM}=\frac{\alpha^2M_h^3}{256\pi^3v^2}\Big|Q^2_{H^{\pm}}\frac{v\mu_{hH^+H^-}}{2M^2_{H^{\pm}}}F_0(\tau_{H^{\pm}})+Q^2_{H_3^{\pm}}\frac{v\mu_{hH_3^+H_3^-}}{2M^2_{H_3^{\pm}}}F_0(\tau_{H_3^{\pm}})+C\Big|,
\end{equation}
where, $C$ is the SM contribution, $ C=\sum_fN_f^cQ_f^2y_fF_{1/2}(\tau_f)+y_WF_1(\tau_W)$ and $\tau_x=\frac{M_h^2}{aM_X^2}$. $Q_i$ denote electric charge of corresponding particles and $N_f^c$ is the color factor. Higgs $h$ coupling to $f\overline{f}$ and $WW$ is denoted by $y_f=y_f^{\rm SM} \frac{\sin\alpha}{\cos\beta}$ and $y_W=y_w^{\rm SM}\cos(\beta-\alpha)$. $\mu_{hH^+H^-}$ and $\mu_{hH_3^+H_3^-}$ stands for corresponding couplings for the $hH_+H^-$ and $hH_3^+H_3^-$ respectively, which are defined below with the loop function $F_{(0,1/2,1)}(\tau)$~\cite{Djouadi:2005gj},
\begin{equation*}
\begin{split}
\mu_{hH^+H^-}=&[2\kappa_4\sin\beta\cos\beta\cos\gamma+\cos\beta^2(\kappa_3v_1\cos\gamma+4\kappa_2v_2\sin\gamma)\nn\\
&+\sin\beta^2(\kappa_4\sin\gamma+\kappa_1v_1\cos\gamma+\kappa_3v_2\sin\gamma)]\approx \kappa_3v_{SM},\\
\mu_{hH_3^+H_3^-}=&\kappa_3^{DM}v_1[\sin\alpha\sin\beta-\cos\alpha\cos\beta],\\
F_0(\tau)=&[\tau-f(\tau)]\tau^{-2},\\
F_{1/2}(\tau)=&2[\tau+(\tau-1)f(\tau)]\tau^{-2},\\
F_1(\tau)=&-[2\tau^2+3\tau+3(2\tau-1)f(\tau)]\tau^{-2},\\
\text{with,}\quad f(\tau)=& \begin{cases} 
(\sin^{-1}\sqrt{\tau})^2, \quad\quad\quad\quad\quad\quad \tau\leq 1,\\ 
-\frac{1}{4}[\ln\frac{1+\sqrt{ 1-\tau^{-1}}}{1-\sqrt{1-\tau^{-1}}}-i\pi]^2\quad\quad \tau>1.
\end{cases}\\
\end{split}
\end{equation*}
\subsection{Bounds from dark matter } 
Various results from the WMAP satellite, combined with other cosmological measurements, we got the constrained dark matter relic density to $\Omega_{DM} h^2=0.1198\pm0.0026$ \cite{Ade:2013zuv}. The dark matter sector of this model is behaving quite similarly as in the normal inert doublet model \cite{Honorez:2010re, Gustafsson:2012aj, Goudelis:2013uca, Arhrib:2013ela}.
However, in the presence of an extra SM type Higgs doublet, these pictures get slightly disturbed. Both the annihilation and co-annihilation are modified, and we find a larger region of allowed parameter spaces in this model than other ID models. 
Within the ID sector, the lightest neutral scalar ($H_3/A_3$) serves as an ID dark matter candidate. In this model, we consider $H_3$ to be a dark matter candidate, and it is to be noted that the region of the dark matter parameter spaces will be slightly changed for the pseudoscalar as a dark matter candidate $A_3$. 
We have used {\tt FeynRules}~\cite{Alloul:2013bka} to construct our model and relic density calculations are carried out using {\tt MicrOMEGAs} \cite{Belanger:2018mqt}. We carry out details discussion about DM mass in the numerical analysis section.

    At present, the most stringent limit on the spin-independent component of elastic
scattering cross section $\sigma^{SI}_{p}<4.1\times 10^{-47}~{\rm cm^2}$ for $M_{DM}\simeq 30$ GeV~\cite{Aprile:2018dbl}. 
In this analysis, we get the dark matter parameter space, which is allowed by direct detection data and other theoretical as well as experimental constraints. 
In this model, we took a minimal Higgs portal coupling, which is allowed by direct detection cross-section and invisible Higgs decay width (for $M_{DM}<M_h/2$) \cite{Duerr:2015aka}.
The dark matter indirect detection constraints such as Fermi-LAT data~\cite{FermiLAT:2011ab}, PAMELA~\cite{Adriani:2010rc} and AMS02~\cite{Aguilar:2013qda,Aguilar:2016vqr} restrict arbitrary Higgs portal couplings~\cite{Karwin:2016tsw, Gaskins:2016cha}.
It is also possible to explain various observations in the indirect DM detection experiments like, combine results from LHC-14 Monojet + XENON1T + AMS02 antiproton flux~\cite{Arhrib:2013ela} from this model. Moreover, we have checked that the dark matter self-annihilation cross-sections for the allowed parameter space. The tiny Higgs portal coupling which is allowed by the direct detection data gives the self-annihilation cross-section, $<\sigma v>\, \lesssim \, \mathcal{O}(10^{-25})~{\rm cm^3s^{-1}}$. The relic density is adjusted by the other (co)annihilation processes and these allowed points do not exceeds the indirect detection bounds. 
However, we do not discuss these bounds here explicitly, as these estimations involve proper knowledge of the astrophysical backgrounds and an assumption of the DM halo profile which contains some arbitrariness.

\subsection{Neutrino mass and mixing angles}
The diagonalize neutrino mass matrix $M_{\nu}$ is achieved as,
\begin{equation}
\label{eq:2}
\text{Diag}(m_{1},m_{2},m_{3})=U_{PMNS} \ M_{\nu}\ U_{PMNS}^{T},
\end{equation}
where, $m_i$(for $i=1,2,3)$ stands for three active neutrino masses. Conventionally the leptonic mixing matrix for active neutrino is parameterized as,
\begin{equation}
U_{PMNS}={\begin{pmatrix}
	c_{12}c_{13}&s_{12}c_{13}&s_{13}e^{-i\delta}\\
	-s_{12}c_{23}-c_{12}s_{23}s_{13}e^{i\delta}&c_{12}c_{23}-s_{12}s_{23}s_{13}e^{i\delta}&s_{23}c_{13}\\
	s_{12}s_{23}-c_{12}c_{23}s_{13}e^{i\delta}&-c_{12}s_{23}-s_{12}c_{23}s_{13}e^{i\delta}&c_{23}c_{13}\\
	\end{pmatrix}}.P.
\end{equation}
The abbreviations used are $c_{ij}=\cos\theta_{ij}$ , $s_{ij}=\sin\theta_{ij}$ where $\theta_{ij}$ stands for active mixing angles with $i,j=1,2,3$ and $P$ would be a unit matrix \textbf{1} in the Dirac case but in Majorana case $P=diag(1~e^{i\alpha}~e^{i(\beta+\delta)})$~\footnote{One should not get confused with the Majorana phases ($\alpha$ and $\beta$) we have used in the fermion sector. They are completely irrelevant to the mixing angles used in the scalar sector. }. The Dirac and Majorana CP-violating phases are simply represented by $\delta$ and ($\alpha, \beta$) in the $U_{PMNS}$ respectively. Since we have included one extra generation of neutrino along with the active neutrinos in our model thus, the final neutrino mixing matrix for the active-sterile mixing takes $4\times4$ form as,
\begin{equation}
V\simeq 
\begin{pmatrix}
(1-\frac{1}{2}RR^{\dagger})U_{PMNS} & R \\ -R^{\dagger}U_{PMNS} & 1-\frac{1}{2}R^{\dagger}R
\end{pmatrix},
\end{equation}
here, $R=M_{D}M_{R}^{-1}M_{S}^{T}(M_{S}M_{R}^{-1}M_{S}^{T})^{-1}$ is a $3\times1$ matrix governed by the strength of the active-sterile mixing, {\it i.e.}, the ratio $\frac{\mathcal{O}(M_D)}{\mathcal{O}(M_S)}$. As a result of new physics, the trivial $3\times3$ unitary leptonic mixing matrix, $U_{PMNS}$ may slightly deviate from it's generic unitarity behaviour~\cite{Antusch:2006vwa,Akhmedov:2013hec}. In general, the mixing between active and sterile lead to non-unitarity in the $U_{PMNS}$ matrix. However, in our study, we have consider a minimal mixing between the active-sterile neutrinos, which doesn't bother the active neutrino scenario. Moreover, current electroweak precision measurements and neutrino oscillation data have constrained $U_{PMNS}$ to be unitary at the $\mathcal{O}(10^{-2})$ level \cite{Xing:2019vks}. 
The eV scaled sterile neutrino can be added to the standard 3-neutrino mass states in NH: $m_1\ll m_2<m_3\ll m_4$ as well as IH: $m_3\ll m_1<m_2\ll m_4$. One may diagonalized light neutrino mass matrix for NH and IH are modified as $m_{\nu}^{NH}=\text {diag}(0, \sqrt{\Delta m_{21}^{2}}, \sqrt{\Delta m_{21}^{2}+\Delta m_{31}^{2}}$,
$\sqrt{\Delta m_{41}^{2}})$ and  $m_{\nu}^{IH}=\text{diag}(\sqrt{\Delta m_{31}^{2}},\sqrt{\Delta m_{21}^{2}+\Delta m_{31}^{2}},0,\sqrt{\Delta m_{43}^{2}})$ respectively. The lightest neutrino mass is zero in both the mass ordering as demanded by the MES framework \cite{Zhang:2011vh}. Here, $\Delta m_{41}^{2}(\Delta m_{43}^{2})$ is the active-sterile mass square difference for NH and IH respectively. Using MES, the mass matrices obtained for active neutrinos and active-sterile mixing elements are shown in table \ref{tab:nh}. It is to be noted that, we have used the light neutrino parameters satisfying the bounds from $\mu\rightarrow e\gamma$ \cite{Dohmen:1993mp}.
\begin{table}[h!] 
	\resizebox{0.99\hsize}{!}{
	\begin{tabular}{|c|c|c|c|c|}
		\hline
		Ordering & Structures & $-m_{\nu}$&$m_S$& $R$ \\
		\hline
		NH & 
		$\begin{aligned}
		&
		M_R=\begin{pmatrix}
		d&0&0\\
		0&e&0\\
		0&0&f\\
		\end{pmatrix}\\
		& M_{D}= \begin{pmatrix}
		b&b&c+p\\
		0&b+p&c\\
		p&b&c\\
		\end{pmatrix}\\
		& M_{S}= \begin{pmatrix}
		g&0&0\\
		\end{pmatrix}\\
		\end{aligned}$
		& $\begin{pmatrix}
		\frac {b^2} {e} + \frac {(c + p)^2} {f} &\frac {b (b + 
			p)} {e} + \frac {c (c + 
			p)} {f} &\frac {b^2} {e} + \frac {c (c + p)} {f} \\
		\frac {b (b + p)} {e} + \frac {c (c + p)} {f} &\frac {(b + 
			p)^2} {e} + \frac {c^2} {f} &\frac {b (b + 
			p)} {e} + \frac {c^2} {f} \\
		\frac {b^2} {e} + \frac {c (c + p)} {f} &\frac {b (b + 
			p)} {e} + \frac {c^2} {f} &\frac {b^2} {e} + \frac {c^2}
		{f} \\
		\end{pmatrix}$&$\simeq \frac{g^2}{10^4}$ & $\simeq{\begin{pmatrix}
			\frac{b}{g}\\ 0\\ \frac{p}{g}\\
			\end{pmatrix}}$\\
		\hline
		IH& 
		$\begin{aligned}
		&
		M_R=\begin{pmatrix}
		d&0&0\\
		0&e&0\\
		0&0&f\\
		\end{pmatrix}\\
		& M_{D}= \begin{pmatrix}
		b&-b&c+p\\
		0&-b+p&c\\
		p&2b&c\\
		\end{pmatrix}\\
		& M_{S}= \begin{pmatrix}
		g&0&0\\
		\end{pmatrix}\\
		\end{aligned}$
		& $\begin{pmatrix}
		\frac {b^2} {e} + \frac {(c + p)^2} {f} &  \frac {b(b - 
			p)} {e} + \frac {c (c + 
			p)} {f} & \frac {-2 b^2} {e} + \frac {c (c + p)} {f} \\
		\frac {b(b - p)} {e} + \frac {c (c + p)} {f} &\frac {(b - 
			p)^2} {e} + \frac {c^2} {f} &\frac {-2 b (b - 
			p)} {e} + \frac {c^2} {f} \\
		-\frac {2 b^2} {e} + \frac {c (c + p)} {f} & - \frac {2 b (b - 
			p)} {e} + \frac {c^2} {f} &\frac {4 b^2} {e} + \frac {c^2}
		{f} \\
		\end{pmatrix}$& $\simeq \frac{g^2}{10^4}$&$\simeq{\begin{pmatrix}
			\frac{b}{g}\\ 0\\ \frac{p}{g}\\
			\end{pmatrix}}$\\
		\hline
	\end{tabular}}
	\caption{The light neutrino mass matrices ($m_{\nu}$), sterile mass ($m_S$) and active-sterile mixing patterns ($R$) with corresponding $M_D$, $M_R$ and $M_S$ matrices for NH and IH mass pattern.  }\label{tab:nh}
\end{table}
\subsection{Baryogenesis via Thermal leptogenesis}
We consider a hierarchical mass pattern for RH neutrinos, among which the lightest will decay to a Higgs doublet and a lepton doublet. This decay would produce sufficient lepton asymmetry to give rise to the observed baryon asymmetry of the Universe. Both baryon number ($B$) and lepton number ($L$) are conserved independently in the SM renormalizable Lagrangian. However, due to chiral anomaly, there are non-perturbative gauge field configurations \cite{Callan:1976je}, which produces the anomalous $B+L$ violation ($B-L$ is already conserved). These whole process of conversion of lepton asymmetry to baryon asymmetry via $B+L$ violation is popularly termed as "sphalerons" \cite{Klinkhamer:1984di}.
We have used the parametrization from \cite{Davidson:2008bu}, where, the working formula of baryon asymmetry produced is given by,
\begin{equation}\label{yb}
Y_B=ck\frac{\epsilon_{11}}{g_{*}}.
\end{equation}
The quantities involved in this equation \ref{yb} can be explained as follows,
\begin{itemize}
	\item $c$ is the factor that measures the fraction of lepton asymmetry that being converted to baryon asymmetry. This value is approximately $12/37$.
	\item $k$ is the dilution factor produced due to wash out processes, which can be parametrized as,
	\begin{equation}\label{sk}
	\begin{split}
	k &\simeq \sqrt{0.1K}exp\big[\frac{-4}{3(0.1K)^{0.25}}\big], \quad \text{for} \quad K\geq10^6,\\
	&\simeq\frac{0.3}{K(lnK)^{0.6}}, \quad \text{for}\quad 10\leq K\leq 10^6,\\
	&\simeq\frac{1}{2\sqrt{K^2+9}}, \quad\text{for} \quad0\leq K \leq 10.\\
	\end{split}
	\end{equation}
Here, $K$ is defined as,
	\begin{equation}
	K=\frac{\Gamma_1}{H(T=M_{\nu_{R1}})}=\frac{(\lambda^{\dagger}\lambda)_{11}M_{\nu_{R1}}}{8\pi}\frac{M_{Planck}}{1.66\sqrt{g_{*}}M^2_{\nu_{R1}}},
	\end{equation}
here, $\Gamma_1$ is the decay width of $\nu_{R1}$, defined as,  $\Gamma_1=\frac{(\lambda^{\dagger}\lambda)_{11}M_{\nu_{R1}}}{8\pi} $ and the Hubble constant at $T=M_{\nu_{R1}}$ is defined as $H(T=M_{\nu_{R1}})=\frac{M_{Planck}}{1.66\sqrt{g_{*}}M^2_{\nu_{R1}}}$. 
\item $g_{*}$ is the massless relativistic degree of freedom in the thermal bath and within SM, it is approximately $110$.
\item $\epsilon_{11}$ is the lepton asymmetry produced by the decay of the lightest RH neutrino $\nu_{R1}$. This can be formulated as below.\\
To produce non-vanishing lepton asymmetry, the decay of $\nu_{R1}$ must have lepton number violating process with different decay rates to a final state with particle and anti-particle. Asymmetry in lepton flavor $\alpha$ produced in the decay of $\nu_{R1}$ is defined as,
	\begin{equation}
	\epsilon_{\alpha\alpha}=\frac{\Gamma(\nu_{R1} \rightarrow l_{\alpha}\phi_i)-\Gamma(\nu_{R1} \rightarrow \overline{l_{\alpha}}\overline{\phi_i})}{\Gamma(\nu_{R1} \rightarrow l \phi_i)+\Gamma(\nu_{R1} \rightarrow \overline{l}\overline{\phi_i})},
	\end{equation}
where, $\overline{l_{\alpha}}$ is the antiparticle of $l_{\alpha}$ and $\phi_i$ is the lightest Higgs doublet present in our model. Following the calculation for non-degenerate RH mass\footnote{For degenerate mass with mass spiting equal to decay width, one has to consider resonant leptogenesis.}, from the work of \cite{Davidson:2008bu}, we obtain the asymmetry term as,
\begin{equation}
	\begin{split}
	\epsilon_{\alpha\alpha}= & \frac{1}{8\pi}\frac{1}{[\lambda^{\dagger}\lambda]_{11}}\sum_{j}^{2,3}\text{Im} {(\lambda_{\alpha 1}^{*})(\lambda^{\dagger}\lambda)_{1j}\lambda_{\alpha j}} g(x_{j})\\
	& + \frac{1}{8\pi}\frac{1}{[\lambda^{\dagger}\lambda]_{11}}\sum_{j}^{2,3}\text{Im} {(\lambda_{\alpha 1}^{*})(\lambda^{\dagger}\lambda)_{1j}\lambda_{\alpha j}} \frac{1}{1-x_j}.
	\end{split}\label{epsi}
\end{equation}
Here, $x_j\equiv \frac{M_j^2}{M_1^2}$ and within the SM $g(x_j)$ is defined as,
	\begin{equation}
	g(x_j)=\sqrt{x_j}\Big(\frac{2-x_j-(1-x_j^2)\text{ln}(1+x_j/x_j)}{1-x_j}\Big).
	\end{equation}
The second line from equation \eqref{epsi} violates the single lepton flavors, however, it conserves the total lepton number, thus it vanishes when we take the sum over $\alpha$,
	\begin{equation}\label{ep}
	\epsilon_{11}\equiv\sum_{\alpha}\epsilon_{\alpha\alpha}=\frac{1}{8\pi}\frac{1}{[\lambda^{\dagger}\lambda]_{11}}\sum_{j}^{2,3}\text{Im}{[(\lambda^{\dagger}\lambda)_{1j}]^2} g(x_{j}).
	\end{equation}
The $\lambda$ used here is the Yukawa matrix generated from the Dirac mass matrix and the corresponding index in the suffix says the position of the matrix element.
\end{itemize} 
Now, baryon asymmetry of the Universe can be calculated from equation \eqref{yb} followed by the evaluation of lepton asymmetry using equation \eqref{ep}. The Yukawa matrix is constructed from the solved model parameters $b,c$ and $p$, which is analogous to the $3\times3$ Dirac mass matrix. Within our study the $K$ value lies within the range $10\leq K\leq 10^6$, hence, we have used the second parametrization of the dilution factor from equation \eqref{sk}.
\section{Numerical analysis}\label{s4}
\subsection{Dark matter}
In this section, we discuss the numerical analysis and new bounds on the DM mass of the model.
As we know, the observed relic density through annihilation in this model mainly relies on the dark matter mass $M_{H_3}$, Higgs (both $h$ and $H$) portal coupling. These couplings are primarily depending upon the coupling $\kappa_L$ and the mixing angles $\alpha$ and $\beta$. The annihilation could also be affected by the mass of the more massive Higgs particle.
If we decrease the mass difference between the LSP and nLSPs (similar to the IDM), co-annihilation channels start to play a crucial role. The mass differences $\Delta M_{A_3}=M_{A_3}-M_{H_3}$ and $\Delta M_{H_3^\pm}=M_{H_3^\pm}-M_{H_3}$  are important here in calculating the relic abundance.
The other $Z_4$-even charged ($H_1^\pm$) and pseudoscalar ($A_1$) particle are also come into this picture depending on their masses and the mixing angles $\alpha$ and $\beta$.

\begin{figure}[h!]
		\includegraphics[width=2.8in,height=2.8in, angle=0]{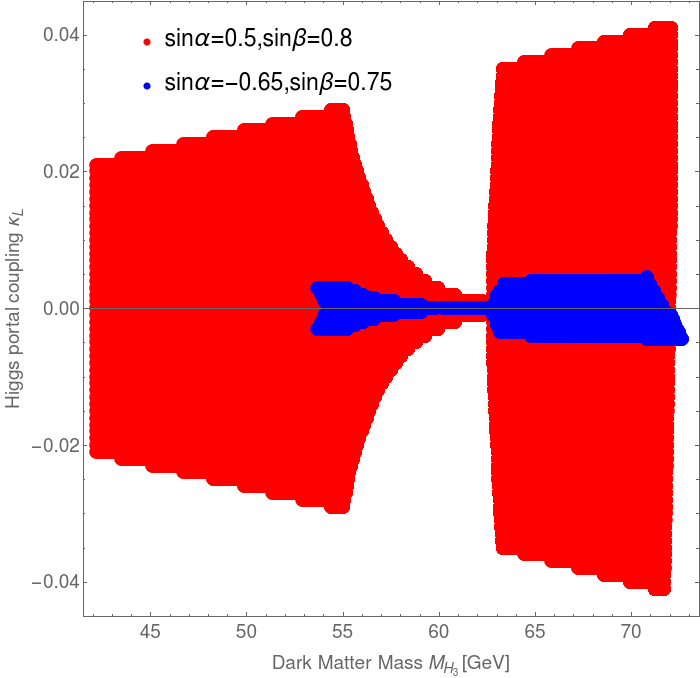}
		\includegraphics[width=2.8in,height=2.8in, angle=0]{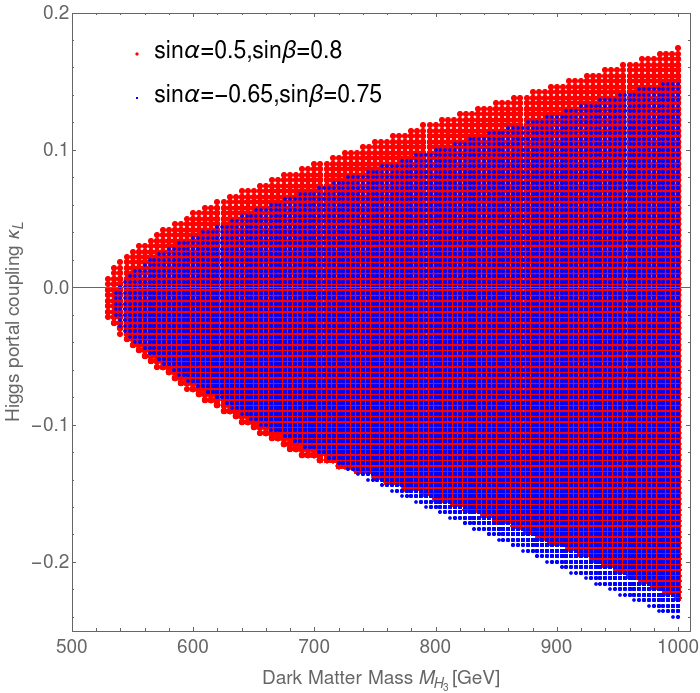}
	\caption{Left plot stands for low mass regime whereas right one indicates high mass regime. The red points correspond to $\cos(\beta-\alpha)\sim 0.92$ and blue $\cos(\beta-\alpha)\sim 0.015$. We keep the heavier Higgs masses fixed at $M_H=400$ GeV and $M_{A,H^\pm}=430$ GeV. These values are allowed by the experimental and theoretical constraints. We varied other scalar masses of the dark matter sector so that we can get relic density via dark matter annihilation, co-annihilation and the combined effect of these.}
	\label{dmplot}
\end{figure}

We performed scans in the four dimensional parameter space. Dark matter mass, $M_{H_3}$ is varied from 5 GeV to 1000 GeV and $\kappa_L$ from $-0.25$ to $0.25$ with a step size $0.001$,  $\Delta M_{A_3}$ from $0$ to $20$ GeV with a step size $0.2$ GeV. We also fixed $\Delta M_{H_3^\pm}$=100 GeV to avoid the collider constraints~\cite{Goodman:2010ku, Djouadi:2011aa}.
The heavier Higgs masses are fixed at $M_H=400$ GeV and $M_{A,H^\pm}=430$ GeV. Two different regimes for a fixed value of the mixing angles  $\alpha$ and $\beta$ are obtained, and we define these as a low and high mass regime. These mass regimes for two different values of $\alpha$ and $\beta$ are shown in Fig.~\ref{dmplot}.
Left plot stands for low mass regime whereas right one indicates high mass regime. The red points correspond to $\cos(\beta-\alpha)\sim 0.92$ and blue $\cos(\beta-\alpha)\sim 0.015$.
It is to be noted that, these points pass through all the experimental as well as theoretical constraints, that we have discussed in the previous section. For the low mass, a dominant part of the points ruled out by the Higgs/Z invisible decay width and direct detection constraints \cite{Aprile:2018dbl}. The STU parameter discards a part of the region, where $\Delta M_{A_3}$ is large, absolute stability of scalar potential and perturbativity, as $\Delta M_{A_3}$ is directly proportional to $\kappa_5^{DM}$. 

In the low mass regime $M_{H_3}<10$ GeV, DM dominantly annihilates into the SM fermions only, and this annihilation cross-section remains small due to the small coupling strength and mass. Hence one would get overabundance as $\Omega_{DM} h^2 \propto \frac{1}{<\sigma v >} $. The abundance in this model is roughly
$1/\cos ^2(\beta-\alpha)$ times larger than the normal IDM abundance~\cite{Khan:2016sxm}.
In particular, within the mass regime $10-73$  GeV, correct values for $\Omega_{DM} h^2$ (within $3\sigma$) will be produced due to the contribution of the DM annihilation, co-annihilation or combined effect of these two processes. While the mass regime $10-42.2$ GeV is ruled out from the constraints of the decay width of SM gauge $W,Z$~\cite{Bertone:2004pz} and/or Higgs $h_1$ bosons ~\cite{Diaz:2015pyv}. The direct detection (DD) processes also played a strong role to ruled out this regime.
It is to be noted that in the normal IDM ~\cite{Datta:2016nfz, Deshpande:1977rw, Arhrib:2013ela, Goudelis:2013uca}, $M_{DM}<54$  GeV regime is ruled out from the similar constraints. In this model, the DM mass $42.2-54$ GeV up to $72.65$ GeV regime is still allowed due to the presence of the other $Z_4$-even scalar particles. We get the larger allowed region in the parameter spaces depending on the mixing angles $\alpha$ and $\beta$, and these regions are explained in Fig.~\ref{dmplot}. A few examples of benchmark points presented in Table.~\ref{BMPlow} and \ref{BMPhigh}. 
\begin{table}
\resizebox{0.98\hsize}{!}{
\begin{tabular}{|c|c|c|c|c|c|c|c|}
\hline
BMP-low &$M_{DM}$ [GeV]&$\kappa_{L}$& $M_{A_3}$ [GeV]&$~\sin\alpha~$&$~\sin\beta~$&Relic density $~\Omega_{DM} h^2~$&DD cross-section [$\rm cm^2$]\\
\hline
\hline
I & 42.20 & $-0.001$ & 53.4 &  0.50 & 0.80 & 0.1266 & $2.24\times 10^{-49}$\\
II& 53.65 & $0.003$ &62.85&-0.65 & 0.75 & 0.1241  & $1.02\times 10^{-46}$\\
\hline
\hline
III&$55.00$&$-0.029$&$73.80$&$0.50$&$0.80$&$0.1121$&$1.12\times 10^{-46}$\\
IV&$55.00$&$0.001$&$63.65$&$-0.65$&$0.75$&$0.1161$&$1.08\times 10^{-47}$\\
\hline
\hline
V&$65.00$&$-0.036$&$74.$&$0.50$&$0.80$&$0.1257$&$1.24\times 10^{-46}$\\
VI&$65.00$&$-0.001$&$73.4$&$-0.65$&$0.75$&$0.1155$&$7.8\times 10^{-48}$\\
\hline
\hline
VII& $72.05$ & $0.041$ &$91.85$ &$0.50$&$0.80$&$0.1152$&$1.31 \times 10^{-46}$\\
VIII&$72.05$&$-0.001$&$93.85$&$-0.65$&$0.75$&$0.1140$&$6.38\times 10^{-48}$\\
\hline
\end{tabular}}
\caption{Benchmark points for low dark matter mass regime. Each horizontal block contain two BMPs, corresponding to different set of mixing angles ($\alpha$ and $\beta$).}
\label{BMPlow}
\end{table}

\begin{figure}[h!]
  \begin{center}
        \includegraphics[width=5in,height=.8in, angle=0]{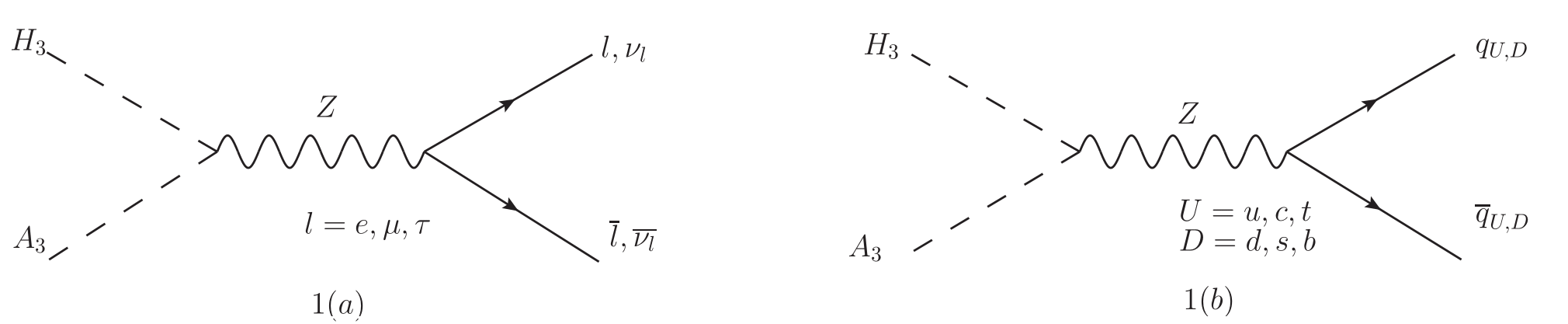}\\
        \includegraphics[width=5in,height=.8in, angle=0]{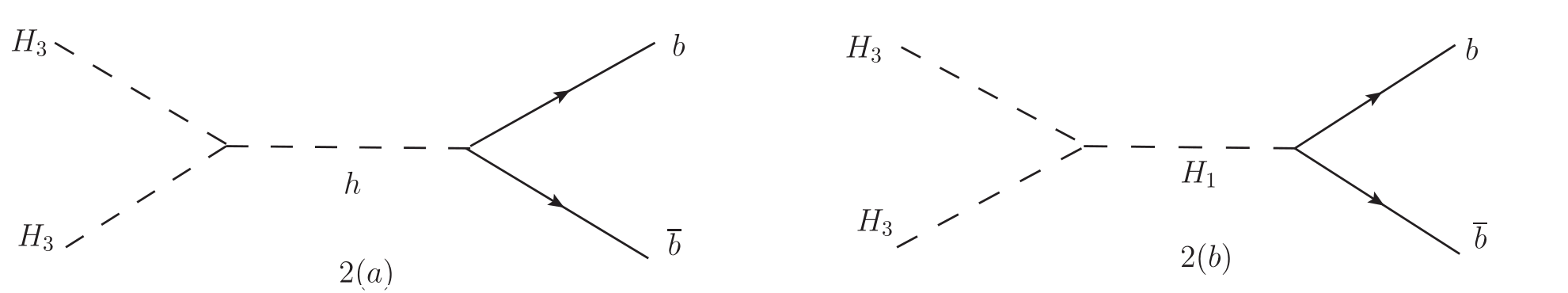}\\
        \includegraphics[width=5in,height=.8in, angle=0]{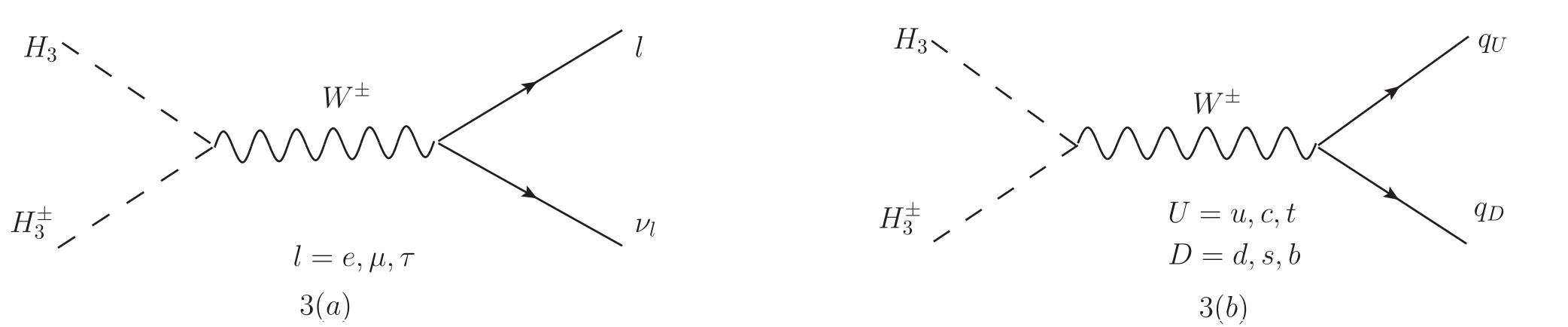}
  \end{center}
\caption{The upper two diagrams stand for the co-annihilation channels $H_3 A_3 \rightarrow Z \rightarrow $ $ XX$, where $X={\rm SM~ quarks/leptons}$, the middles diagrams are $H_3 H_3^\pm \rightarrow W^\pm \rightarrow XX$. The dominant annihilation diagrams are $H_3 H_3 \rightarrow h/H \rightarrow {\rm b\bar{b}}$. These channels are effective and/or dominant for the dark matter mass regions $10-70$ GeV regime.}
    \label{Diag1}
\end{figure}

In the table~\ref{BMPlow}, few benchmark points
have presented for low dark matter mass regime.
The first BMP-I corresponds to dark matter mass $42.2$ GeV with Higgs portal couplings $\kappa_L \cos(\beta+\alpha)= 0.12 \kappa_L$ ($\sin\alpha=0.5$ and $\sin\beta=0.8$) and $M_{A_3}=53.4$ GeV. As the $\Delta M_{A_3}$ ($=11.2$ GeV) is small, we get the relic density mainly dominated by the $Z$-mediated co-annihilation channels $H_3 A_3\rightarrow Z \rightarrow XX$, where $X={\rm SM~ fermions}$. One can find the corresponding diagrams (upper two) in Fig.~\ref{Diag1}. Here, $M_{A_3}+M_{H_3}=95.6>M_Z$, hence this process is allowed by the $Z$-boson invisible decay width constraints. On the other hand, the Higgs portal coupling is also very small $\sim 10^{-3}$; hence, this point is also allowed by the invisible Higgs decay width and direct detection constraints. It is also to be noted that, in the normal inert doublet model, one may get the exact relic density for the DM mass below $54$ GeV, but one of these constraints will restrict this point. For our model, this may be considered as a new finding, as this small DM mass not been discussed in the literature in detail.

\begin{figure}
  \begin{center}
        \includegraphics[width=5in,height=.8in, angle=0]{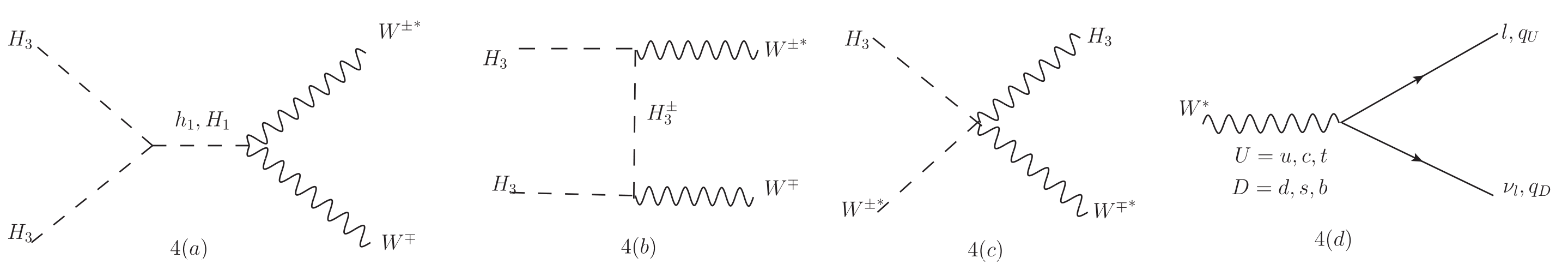}
        \includegraphics[width=5in,height=.8in, angle=0]{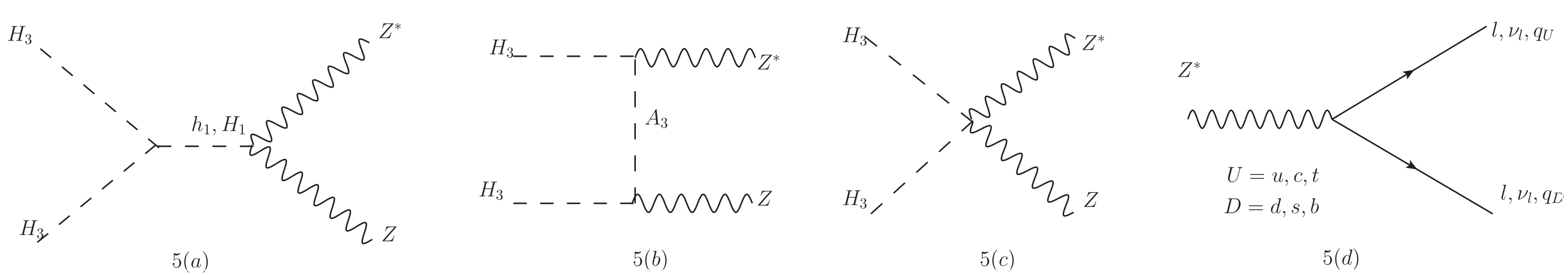}
  \end{center}
  \caption{These channels are effective and dominant for the dark matter mass regions $>70$ GeV regime. For the dark matter mass $M_{DM}<M_W / M_Z$, gauge bosons further decaying into SM fermions. The relic density is dominated by the $H_3 H_3 \rightarrow V V^*$, $ V^* \rightarrow {\rm SM~ quarks/leptons} $, where $V=W,Z$. For the high mass regime the dominant processes are $H_3 H_3 \rightarrow V V$.}
    \label{Diag2}
\end{figure}

We changed the mixing angles $\sin\alpha=-0.65$ and $\sin\beta=0.75$, thus the Higgs portal coupling becomes $0.98 \kappa_L$. The dark matter masses $M_{H_3}=42.2$ GeV for this mixing angle get constrained by both the Higgs invisible decay width and direct detection cross-section. 
We get the allowed point for the next minimum dark matter mass $M_{H_3}=53.65$ GeV, for the Higgs coupling $0.98\kappa_L$. It could be understood as follows: as we change these angles, the Higgs portal coupling becomes too large, which violates the Higgs invisible decay width~\cite{Khachatryan:2016whc} and the direct detection cross-section bounds for the dark matter mass $42.2$ GeV. Hence, we increase the mass to get the allowed relic density. 
Here, $Z$-mediated  co-annihilation channels are contributing $78\%$ of the total processes and dominates the whole effective annihilation process. The annihilation processes $H_3 H_3 \rightarrow b\bar{b} (16\%)$ and $H_3 H_3 \rightarrow W^\pm W^{\mp *}(5\%)$ also played a significant role to achieve the exact relic density. All the diagrams in Fig.~\ref{Diag1} ( upper and middle ) and Fig. \ref{Diag2} (upper) are relevant here. It is to be noted that the processes  $H_3 H_3^\pm \rightarrow W^\pm \rightarrow XX$, where $X={\rm SM~ quarks/leptons}$ are not important in our case as we consider $\Delta M_{H_3^\pm}$=100 GeV.
For BMP-III and IV, exact relic density is obtained with DM mass 55 GeV. The prime difference between these two BMPs arises due to the mixing angles, hence the Higgs portal coupling.
In the first case, $\cos(\alpha+\beta)=0.12$ and $\Delta M_{A_3}=18.80$ GeV with a quite large negative $\kappa_{L}=-0.029$. One can also get allowed relic density for similar positive $\kappa_{L}$ values and it can be understood from Fig.~\ref{dmplot}. In this case, 
the exact relic density is obtained due to the Higgs mediated $H_3 H_3 \rightarrow b\bar{b}(88\%)$ annihilation channel. The other annihilation processes are $H_3 H_3 \rightarrow c\bar{c}, \tau^+\tau^-, WW^*(12\%)$. If we move towards the smaller values of $\kappa_{L}$, the co-annihilation channels become important to get the relic density. For the large values of  $\kappa_{L}$, the Higgs mass resonance region ($55-63$ GeV) of the dark matter are ruled out by one of the constraints as discuss earlier. For example, the dark matter mass $60$ GeV with $\kappa_{L}=\pm 0.01$, the relic density become $\Omega_{DM} h^2=0.01$.\\
For the BMP-IV, very small values of $\kappa_{L}$ are allowed as the total Higgs portal coupling is reduced by the mixing angles. We get exact relic density for 
$\Delta M_{A_3}=8.65$ GeV,
mainly through the co-annihilation channels $H_3 A_3 \rightarrow {\rm SM ~fermions} (92\%)$ which dominates over the DM annihilation channel $H_3 H_3 \rightarrow W^\pm W^{\mp *}$, (8\%). $|\kappa_{L}|>0.003$ region are ruled out by direct detection.
A similar analogy also works for $M_{DM}=65$ GeV.

For BMP-VII and VIII, $M_{DM}=72.05$ GeV, noticeable results are observed for two different mixing angles.
Since the DM mass is close to $W$-boson mass, which allowed to dominate via annihilation channel of $DM,DM\rightarrow W^\pm W^{\mp *}$ ($\sim 77\%$) for small Higgs portal coupling. On the other hand, for larger $\cos(\alpha+\beta)$, co-annihilation processes continues to dominates with $H_3 A_3 \rightarrow {\rm SM ~fermions}$, with a little contribution from dark matter annihilation process into $b\bar{b}$ ($\sim 3\%$) to give rise to the exact relic density. It is to be noted that the effective annihilation cross-section for the DM mass $72.65$ GeV, mixing angles $\sin\alpha=-0.65$ and $\sin\beta=0.75$ become large, hence, we got an under abundance. A negative value of $\kappa_{L}\sim -0.0045$ reduces the effective annihilation cross-section; hence, the exact relic density is observed. However, $\kappa_{L}< -0.0045$ region is ruled out by the direct detection cross-section. The region $72.65$ GeV to $536$ GeV is ruled out as the annihilation rates $H_3 H_3 \rightarrow W^\pm W^\mp, ZZ$ (see Fig.~\ref{Diag2}) are very high, which reduces the relic abundance $\Omega_{DM} h^2 < 0.01$. The negative values of $\kappa_{L}$ may give the exact relic density. However, it will be ruled out the direct detection \cite{LopezHonorez:2010tb}. If we consider, minimal values of $\sin\beta$, one may get the allowed relic abundance, but these regions are again discarded by the constraints of the absolute stability and/or unitarity, perturbativity.\\
\begin{table}
\resizebox{0.98\hsize}{!}{
\begin{tabular}{|c|c|c|c|c|c|c|c|}
\hline
BMP-high &$M_{DM}$ [GeV]&$\kappa_{L}$& $M_{A_3}$ [GeV]&$~\sin\alpha~$&$~\sin\beta~$&Relic density $~\Omega_{DM} h^2~$&DD cross-section [$\rm cm^2$]\\
\hline
\hline
I & 536 & $-0.02$ & 537 &  0.50 & 0.80 & 0.1127 & $5.783\times 10^{-49}$\\
II & 530 & $-0.022$ & 530.5 &  -0.65 & 0.75 & 0.112 & $2.24\times 10^{-49}$\\
\hline
III & 760 & $0.044$ & 761 &  0.5 & 0.80 & 0.1195 & $1.39\times 10^{-48}$\\
IV & 760 & $0.046$ & 765 &  -0.65 & 0.75 & 0.1122 & $1.24\times 10^{-46}$\\
\hline
\end{tabular}}
\caption{Benchmark points for high dark matter mass regime. Each horizontal block contain two BMPs, corresponding to different set of mixing angles ($\alpha$ and $\beta$).}
\label{BMPhigh}
\end{table}
Beyond 536 GeV, we obtain the points by satisfying the constrains discussed in previous subsections. For the high mass region, the DM annihilation channel almost equally contributes, and they get partially canceled out between various diagrams in the limit $M_{DM}\gg M_W$. Usually, $s- channel$ and $p- channel$ mediated processes like $DM,DM\rightarrow W^{\pm}W^{\mp}, ZZ$ get partially cancelled out by $u-channel$ and $t-channel$ processes to give rise to the correct relic bound.  One can find that the sum of the amplitude for these diagrams is proportional to $M_{H_3^{\pm}}^2-M^2_{DM}$~\cite{Khan:2015ipa}, hence, for high DM mass range such cancellation occurs for very small mass difference around 8 GeV. For a different set of $\alpha$ and $\beta$, DM mass vs. Higgs coupling for up to 1000 GeV mass for DM are shown in Fig.~\ref{dmplot}. A slight shift in the allowed regions for two different sets of  $\alpha$ and $\beta$ values is observed. Few BMPs are also shown in table ~\ref{BMPhigh} for the high mass region. 

For the small Higgs coupling strength $0.12\kappa_{L}$, we get the satisfied relic density value with DM mass starting from 536 GeV, while for large $0.98\kappa_{L}$ the starting value for DM mass stands on 530 GeV. In the both the processes, charge scalar decay to the $W$ boson ($H_3^+, H_3^-\rightarrow W^{\pm}W^{\mp}$) are contributing around $16\%$ to $17\%$. However, DM annihilation processes like  $DM,DM\rightarrow W^{\pm}W^{\mp}$ ($\sim 13\%$ for BMP-I and $\sim15\%$ for BMP-II ) and $DM,DM\rightarrow ZZ$ ($\sim10\%$ for BMP-I and $\sim12\%$ for BMP-II) are contributing almost in equal amount, and we get desired range for relic density by partial cancellation among themselves. As we keep increasing the DM mass, the DM annihilation processes are contributing to modest fashion. For BMP-III and BMP-IV, we can find that annihilation processes like $DM,DM\rightarrow W^{\pm}W^{\mp}$($\sim16\%$ form BMP-III and $8\%$ for BMP-IV) and $DM,DM\rightarrow ZZ$ ($\sim9\%$ for BMP-III and $\sim11\%$ for BMP-IV) are contributing to achieving correct relic density. Interestingly, major co-annihilation channels are observed for $M_{DM}=760$ GeV, like $H_3^+H_3^-\rightarrow W^+W^-$. However, their contributions also get suppressed due to the partial cancellation among various diagrams. 
\subsection{Neutrino and Baryogenesis}
In this work, along with the scalar sector, we also tried to shed light on the active neutrino sector in the presence of an extra Higgs doublet, which also takes part in the Lagrangian \eqref{lag} to give mass to the active neutrinos like the SM Higgs. According to our model, both the Higgs doublets acquiring different VEVs. It is to be noted that we consider two different sets of $\sin\beta$. While there is very small distinction between these doublet VEVs for the two sets we have considered ($\sin\beta=0.8$ and $\sin\beta=0.75$), hence no significant difference can be found in neutrino sector for nearly identical $\tan\beta (\sim1.33)$. As the involvement of the Higgs mass can be visualized through the Yukawa coupling of the Lagrangian, we tried to incorporate the result concerning the Yukawa couplings.
\begin{figure}[h!]
		\includegraphics[width=2.8in,height=2.8in, angle=0]{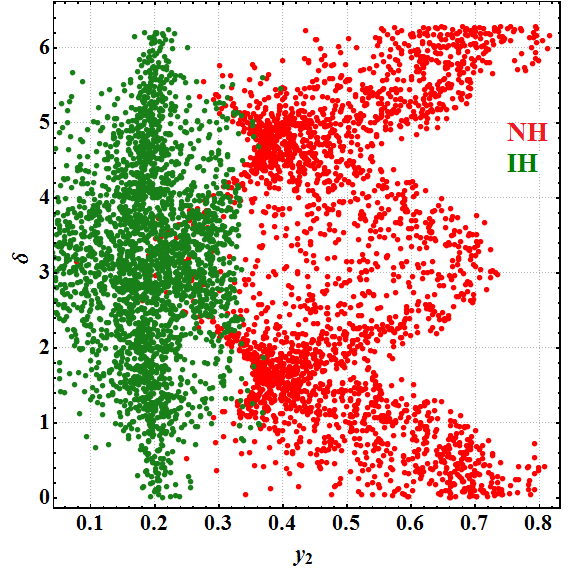}
		\includegraphics[width=2.8in,height=2.8in, angle=0]{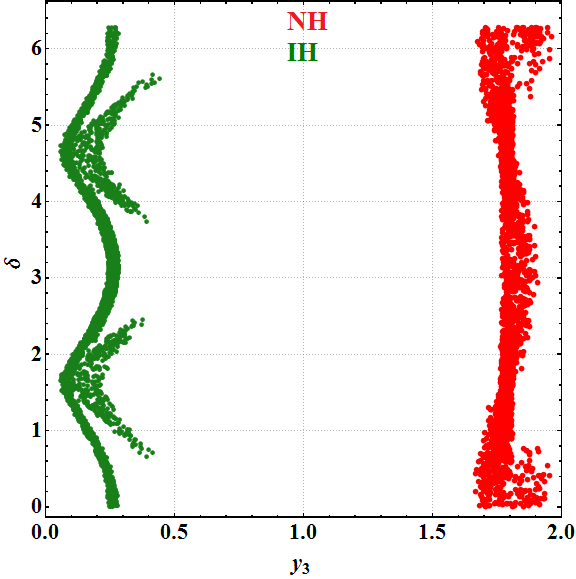}
	\caption{Constrained region in Dirac CP-phase can be seen with the Yukawa coupling. Red dots represents the normal hierarchy while green dots represents inverted hierarchy.} \label{ydel}
\end{figure}
\begin{figure}[h!]
		\includegraphics[width=2.8in,height=2.8in, angle=0]{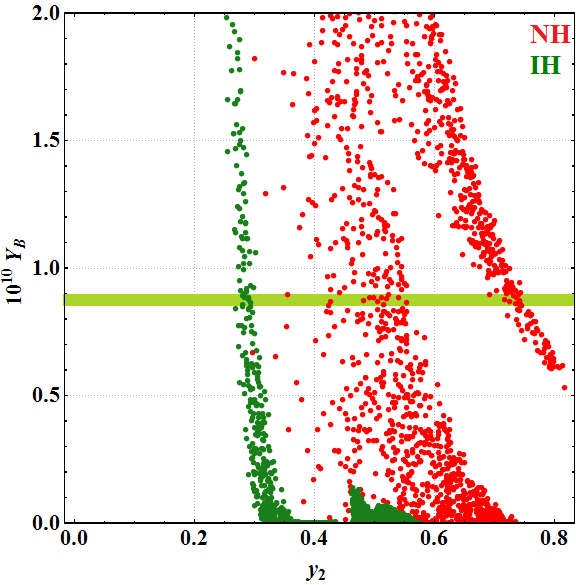}
		\includegraphics[width=2.8in,height=2.8in, angle=0]{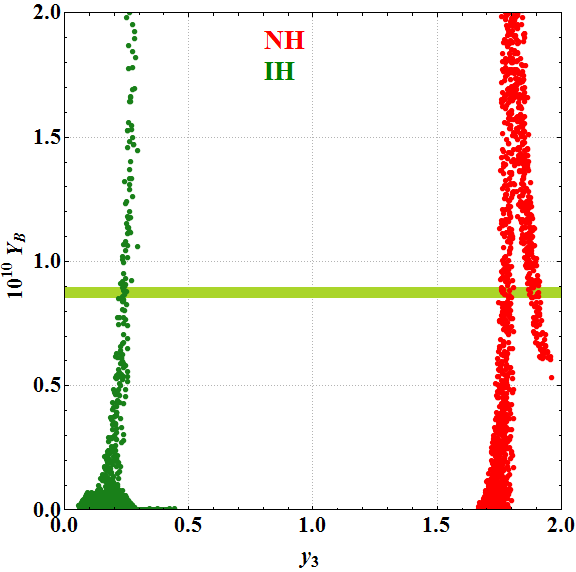}
	\caption{variation of Yukawa coupling with the baryon asymmetry of the Universe. Red dots shows NH and black dots represents IH. The greenish band give the current BAU bound, which is $8.75\pm0.23$.}\label{ybau}
\end{figure}

Here, we present the baryogenesis, including the neutrino mass and mixing angle constraints. Theoretical approaches for light neutrino mass generation and baryogenesis via the mechanism of thermal leptogenesis are discussed in the previous section (see equation~\eqref{amass}, \eqref{yb} respectively). To achieve the observed bound on the baryon asymmetry of the Universe, the Yukawa matrix must be non-zero and complex. 
After solving the model parameters using the latest global fit 3$\sigma$ values of the light neutrino parameters, we can construct the Yukawa matrix. 
Normal and inverted hierarchy are studied simultaneously in the model, and their results are discussed in this work. 
We found interesting bounds on Yukawa coupling with the Dirac CP-phase and the baryon asymmetry of the Universe. 
Even though all the light neutrino parameters along with the Higgs VEV depends upon the Yukawa couplings, the significant contributions can be observed in the reactor mixing angle ($\theta_{13}$) and all the mass squared differences.  

Here, $y_2$ corresponds to the SM Higgs while $y_3$ corresponds to the second Higgs doublet (see eq.~\ref{lag}). 
Major constrains on experimental parameters like $\theta_{13}$ and $m_2$ are coming from $y_2$, which can be seen in the first plot of fig.~\ref{ydel} and \ref{ybau}.
Meanwhile, in the case of $y_3$, along with $\theta_{13}$ and $m_2$, drastic constraints are observed in $m_3$ as well. These dependencies reflect on the second plot of both the fig.~\ref{ydel} and ~\ref{ybau}. The Red dots represent the  NH, and green dots represent IH in both the plots.
In Fig.~\ref{ydel}, variation of Yukawa couplings ($y_2,y_3$) are shown with the Dirac CP-phase.  We found that $y_2$ gets constrained between 0.3-0.7 in NH while for IH large numbers of points are accumulated in between 0.1 and 0.3. Similarly for $y_3$, NH mode is spread around 1.70-1.9 and for IH, its value lie within 0.1-0.5, due to the vanishing lightest neutrino mass ($m_3$). In current dim-5 situation, the Yukawa coupling of  $\mathcal{O}(10^{-2}-1)$~\cite{Abada:2007ux, Ibarra:2010xw, Das:2012ze, Das:2017nvm} are in acceptable range, however, within our study, the large and small values of $y_3$ violates the experimental range of $\Delta m_{31}^2$. For example, large $y_3(\ge2.0)$, $\Delta m_{31}^2$ value exceed the current upper bound of $3\sigma$ value, whereas small $y_3(\le 0.2)$, $\Delta m_{31}^2$ value goes beneath the lower $3\sigma$ bound. 

As BAU value is highly sensitive to the experimental results, very narrow regions are observed with $y_2$ and $y_3$, which  are shown in Fig.~\ref{ybau}. Results corresponding to the Yukawa couplings and BAU are shown in fig.~\ref{ybau}, which verify the successful execution of BAU within the MES framework for both the mass orderings. Large excluded regions in fig. \ref{ybau} are due to the bounds on light neutrino parameters imposed by the Yukawa matrix involved in the baryogenesis calculation. Similar analogy from fig. \ref{ydel}, regarding the bounds on $y_2$ and $y_3$ also works in fig. \ref{ybau}.
\section{Conclusion}\label{s5}
In this paper, we explored a $A_4$ based flavor model along with $Z_4$ discrete symmetry to establish tiny active neutrino mass. Along with this, the generation of non-zero reactor mixing angle ($\theta_{13}$) and simultaneously carried out multi Higgs doublet framework where one of the lightest odd particles behaves as DM candidate. This work is an extension of our previous work on active-sterile phenomenology~\cite{Das:2018qyt}. Hence, we do not carry out the sterile neutrino part in this work. Apart from neutrino phenomenology, the scalar sector is also discussed in great detail. Three sets of SM like Higgs doublets are considered, where two of them acquire some VEV after EWSB and take part in the fermion sector, in particular, they involved in tiny neutrino mass generation. On the other hand, the third Higgs doublet does not acquire any VEV due to the additional $Z_4$ symmetry. 
As a result, the lightest odd particle becomes a viable candidate for dark matter in our model.

In this minimal extended version of the type-I seesaw, we successfully achieved the non-zero $\theta_{13}$ by adding a perturbation in the Dirac mass matrix for both the mass orderings.
The involvement of the high scaled VEVs of $A_4$ singlet flavons $\xi$ and $\xi^{\prime}$ ensure the $B-L$ breaking within our framework, which motivates us to study baryon asymmetry of the Universe within this framework. As the RH masses are considered in non-degenerate fashion, we have successfully able to produce desired lepton asymmetry (with anomalous violation of $B+L$ due to chiral anomaly), which eventually converted to baryon asymmetry by the $sphaleron$ process.

The influence of the Higgs doublets ($\phi_1,\phi_2$) can be seen both in fermion as well as the scalar sector. In the fermion sector,  as the involvement of the Higgs doublets are related to the model parameters via the Yukawa couplings. Relations between the constrained Dirac CP-phase and satisfied baryogenesis results for two different Yukawa couplings are shown, which related via two Higgs' VEV. On the other hand, in the case of the scalar sector, a large and new DM mass region in the parameter spaces is obtained due to the presence of other heavy particles. 
We are successfully able to stretch down the  DM mass at limit up to 42.2 GeV, satisfying all current bounds from various constraints. 
\section{Note added}
A similar analysis on three Higgs (2 active+1 Inert) doublet model has been appeared in Ref.~\cite{Merchand:2019bod}. However, in this work we have studied both the scalar as well as extended fermion sector for the same model considered.

\section{Acknowledgement}
The research work of P.D. and  M.K.D. is supported by the Department of Science and Technology, Government of India under the project grant EMR/2017/001436. The work of  N.K. is partially supported by the Department of Science and Technology, Government of India for SERB-Grant PDF/2017/00372. P.D. would like to thank Biplob Bhattacherjee for the invitation and facility cum hospitality provided at Centre for High Energy Physics (CHEP), IISc Bangalore. 
 \appendix
 \allowdisplaybreaks
 \section{$A_4$ group and product rules}\label{a4p}
 $A_4$, the symmetry group of a tetrahedron, is a discrete non-Abelian group of even permutations of four objects. It has 12 elements with four irreducible representations: three one-dimensional and one three-dimensional which are denoted by $\bf{1}, \bf{1'}, \bf{1''}$ and $\bf{3}$ respectively. $A_4$ can be generated by two basic permutations $S$ and $T$ given by $S=(4321)$ and $T=(2314)$ (For a generic (1234) permutation). One can check immediately as,
 $$S^2=T^3=(ST)^3=1.$$
 The irreducible representations for the $S$ and $T$ basis are different from each other. We have considered the $T$ diagonal basis as the charged lepton mass matrix is diagonal in our case. Their product rules are given as~\cite{Ishimori:2010au},
\begin{equation*}
\begin{split}
 &\bf{1} \otimes \bf{1} = \bf{1}; \bf{1'}\otimes \bf{1'} = \bf{1''}; \bf{1'} \otimes \bf{1''} = \bf{1} ; \bf{1''} \otimes \bf{1''} = \bf{1'}\\
 &\bf{3} \otimes \bf{3} = \bf{1} \otimes \bf{1'} \otimes \bf{1''} \otimes \bf{3}_a \otimes \bf{3}_s 
 \end{split}
\end{equation*}
 where $a$ and $s$ in the subscript corresponds to anti-symmetric and symmetric parts respectively. Denoting two triplets as $(a_1, a_2, a_3)$ and $(b_1, b_2, b_3)$ respectively, their direct product can be decomposed into the direct sum mentioned above as~\cite{Altarelli:2005yp},
 \begin{equation}
 \begin{split}\label{a4r}
 & \bf{1} \backsim a_1b_1+a_2b_3+a_3b_2\\
 & \bf{1'} \backsim a_3b_3+a_2b_1+a_1b_2\\
 & \bf{1''} \backsim a_2b_2+a_1b_3+a_3b_1\\
 &\bf{3}_s \backsim (2a_1b_1-a_2b_3-a_3b_2, 2a_3b_3-a_1b_2-a_2b_1, 2a_2b_2-a_1b_3-a_3b_1)\\
 & \bf{3}_a \backsim (a_2b_3-a_3b_2, a_1b_2-a_2b_1, a_3b_1-a_1b_3)\\
 \end{split}
 \end{equation}
 \section{The mass matrices}\label{matrix}
Following the Lagrangian from equation \eqref{lag}, the respective charged lepton, Dirac, Majorana and sterile mass matrix after acquiring the VEVs looks like,
 \begin{equation}
 M_{l} = \frac{\langle \phi_1\rangle v_m}{\Lambda}\begin{pmatrix}
 y_e&0&0\\
 0&y_{\mu}&0\\
 0&0&y_{\tau}\\
 \end{pmatrix},\
 M^{\prime}_{D}=
 \begin{pmatrix}
 b&b&c\\
 0&b&c\\
 0&b&c\\
 \end{pmatrix},\ M_{R}=\begin{pmatrix}
 d&0&0\\
 0&e&0\\
 0&0&f\\
 \end{pmatrix}, \
 M_{S}= \begin{pmatrix}
 g&0&0\\
 \end{pmatrix}.
 \label{emd}
 \end{equation}
 Here, $b=\frac{\langle \phi_2\rangle v_m}{\Lambda}y_{2} $ and $c=\frac{\langle \phi_3\rangle v_m}{\Lambda}y_{3}$. Other elements are defined as $d=\lambda_{1}v_m, e=\lambda_{2}v_m$, $f=\lambda_{3}v_m$ and $g=\rho v_{\chi}$. \\
 Considering these structure, the light neutrino mass matrix generated from eq.~\ref{amass} takes a symmetric form as,
 \begin{equation}\label{symm}
 m_{\nu}= \begin{pmatrix}
 -\frac {b^2} {e}-\frac{c^2}{f} &-\frac {b^2} {e}-\frac{c^2}{f} &-\frac {b^2} {e}-\frac{c^2}{f} \\
 -\frac {b^2} {e}-\frac{c^2}{f} &-\frac {b^2} {e}-\frac{c^2}{f}&-\frac {b^2} {e}-\frac{c^2}{f} \\
 -\frac {b^2} {e}-\frac{c^2}{f}&-\frac {b^2} {e}-\frac{c^2}{f}&-\frac {b^2} {e}-\frac{c^2}{f}\\
 \end{pmatrix}.
 \end{equation}
This active mass matrix $m_{\nu}$  is a democratic matrix (here we have used $M_D^{\prime}$ in lieu of $M_D$ \eqref{amass}). Only one mixing angle and one mass squared difference can be achieved from it. To generate two mass squared differences and three mixing angles this symmetry must be broken. 
We impose an extra perturbation to the Dirac mass matrix in order to generate non-zero $\theta_{13}$ by introducing $\mu-\tau$ asymmetry in the active mass matrix. New $SU(2)$ singlet flavon fields ($\zeta^{\prime}$ and $\varphi^{\prime}$) are considered and supposed to take $A_4\times Z_4$ charges as same as $\zeta$ and $\varphi$ respectively. After breaking flavor symmetry, they acquire VEV along $\langle\zeta^{\prime}\rangle=(v_p,0,0)$ and $\langle\varphi^{\prime}\rangle=(0,v_p,0)$ directions. Scale of these VEV ($v_p$) in comparison to earlier flavon's VEV ($v_m$) are differ by an order of magnitude ($v_m>v_p$). The Lagrangian that generate the matrix \eqref{pmatrix} can be written as,
 \begin{equation}
 \mathcal{L}_{\mathcal{M_P}} =\frac{y_{1}}{\Lambda}(\overline{l}\tilde{\phi_1}\zeta^{\prime})_{1}\nu_{R1}+\frac{y_{1}}{\Lambda}(\overline{l}\tilde{\phi_1}\varphi^{\prime})_{1^{\prime\prime}}\nu_{R2}+\frac{y_{1}}{\Lambda}(\overline{l}\tilde{\phi_2}\varphi^{\prime})_{1}\nu_{R3}.
 \end{equation}
Hence, the perturbed matrix looks like,
\begin{equation}\label{pmatrix}
 M_{P}=
 \begin{pmatrix}
 0&0&p\\
 0&p&0\\
 p&0&0\\
 \end{pmatrix}.
 \end{equation}
 Hence $M_D$ from eq. \eqref{emd} will take new structure as,
 \begin{equation} \label{md}
 M_D=M^{\prime}_D+M_P=
 \begin{pmatrix}
 b&b&c+p\\
 0&b+p&c\\
 p&b&c\\
 \end{pmatrix}.
 \end{equation}
We also modify the Lagrangian for the $M_D$ matrix by introducing a new triplet flavon $\varphi^{\prime\prime}$ with VEV alignment as $\langle\varphi^{\prime\prime}\rangle \sim (2v_m,-v_m,-v_m)$, which affects only the Dirac neutrino mass matrix and give desirable active-sterile mixing in IH \cite{Zhang:2011vh, Das:2018qyt}. The invariant Yukawa Lagrangian for the $M_D$ matrix in IH mode will be,
 \begin{equation}
 \mathcal{L}_{\mathcal{M_D}}= \frac{y_{2}}{\Lambda}(\overline{l}\tilde{\phi_1}\zeta)_{1}\nu_{R1}+\frac{y_{2}}{\Lambda}(\overline{l}\tilde{\phi_1}\varphi^{\prime\prime})_{1^{\prime\prime}}\nu_{R2}+\frac{y_{3}}{\Lambda}(\overline{l}\tilde{\phi_2}\varphi)_{1}\nu_{R3}.
 \end{equation}
The Dirac mass matrix within IH mode (with perturbation matrix $M_P$) takes new structure as, 
 \begin{equation}
 M_{D}=
 \begin{pmatrix}
 b&-b&c+p\\
 0&-b+p&c\\
 p&2b&c\\
 \end{pmatrix}.
 \end{equation}
\bibliographystyle{utphys}
\bibliography{tevportalnew}

\end{document}